\documentclass{aa}  

\usepackage{graphicx}
\usepackage{txfonts}
\usepackage{color}
\usepackage{xcolor}
\usepackage{upgreek} \newcommand\micron{\ensuremath{\rm \upmu m}\@\xspace} 

\begin{document}

   \title{Pulsation-induced atmospheric dynamics in M-type AGB stars}

   \subtitle{Effects on wind properties, photometric variations and near-IR CO line profiles}

   \author{S. Liljegren
          \inst{1},
          S. H\"ofner\inst{1},
          K. Eriksson\inst{1},
          and
          W. Nowotny\inst{2}
          }

\institute{Division of Astronomy and Space Physics, Department of Physics and Astronomy, Uppsala University, Box 516, SE-751 20 Uppsala, Sweden \\
\email{sofie.liljegren@physics.uu.se}
\and
University of Vienna, Department of Astrophysics, T\"urkenschanzstrasse 17, 1180 Wien, Austria}

   \date{...}

 
  \abstract
{Wind-driving in asymptotic giant branch (AGB) stars is commonly attributed to a two-step process. First, matter in the stellar atmosphere is levitated by shock waves, induced by stellar pulsation, and second, this matter is accelerated by radiation pressure on dust, resulting in a wind. In dynamical atmosphere and wind models the effects of the stellar pulsation are often simulated by a simplistic prescription at the inner boundary.}
{We test a sample of dynamical models for M-type AGB stars, for which we kept the stellar parameters fixed to values characteristic of a typical Mira variable but varied the inner boundary condition. The aim was to evaluate the effect on the resulting atmosphere structure and wind properties. The results of the models are compared to observed mass-loss rates and wind velocities, photometry, and radial velocity curves, and to results from 1D radial pulsation models. The goal is to find boundary conditions which give realistic atmosphere and wind properties.}
{Dynamical atmosphere models are calculated, using the DARWIN code for different combinations of photospheric velocities and luminosity variations. The inner boundary is changed by introducing an offset between maximum expansion of the stellar surface and the luminosity and/or by using an asymmetric shape for the luminosity variation. Ninety-nine different combinations of theses two changes are tested.}
{The model atmospheres are very sensitive to the inner boundary. 
Models that resulted in realistic wind velocities and mass-loss rates, when compared to observations, also produced realistic photometric variations. 
For the models to also reproduce the characteristic radial velocity curve present in Mira stars (derived from CO $\Delta v = 3$ lines), an overall phase shift of 0.2 between the maxima of the luminosity and radial variation had to be introduced. This is a larger phase shift than is found by 1D radial pulsation models.}
{We find that a group of models with different boundary conditions (29 models, including the model with standard boundary conditions) results in realistic velocities and mass-loss rates, and in photometric variations. To achieve the correct line splitting time variation a phase shift is needed. }


   \maketitle
%
%

\section{Introduction}

The mass loss of asymptotic giant branch (AGB) stars through a slow stellar wind is presumably caused by a two-step process. 
First, stellar pulsations create shock waves in the surface layers of the star that levitate matter in the atmosphere. 
This compressed, levitated material then reaches temperatures that are cool enough for dust condensation to occur. 
In this second stage the dust is accelerated outwards by radiation, through scattering or absorption, depending on the chemical composition and size of the dust grains. 
A stellar wind is triggered, as momentum is transferred from the dust to the gas through collisions.

This scenario has been investigated extensively, and is supported by various observations. 
The gas dynamics in the region where shock waves develop are studied through high-resolution spectroscopy, e.g. vibration-rotation lines of CO. 
The CO molecule is stable throughout the atmosphere and the wind acceleration region, making CO lines suitable to probe the inner regions, with regular pulsation, the shock waves and the steady outflow
\citep[see e.g.][]{hinkle_time_1982,nowotny_atmospheric_2005-1,nowotny_atmospheric_2005,nowotny_line_2010}. 
Interferometry and high angular resolution imagining are also becoming important for studying both the structure and the dynamics of AGB star atmospheres, giving new insights into the shock propagation and dust condensation distances \citep[see e.g.][]{chandler_asymmetries_2007,karovicova_new_2013,ohnaka_spatially_2012,ohnaka_clumpy_2016}.

The wind driving process is usually studied with dynamical models \citep[see][ for a recent review of these models]{hof15} which simulate the stellar atmosphere where radiative effects dominate. 
This is a vastly different region from the stellar interior, where convection is the main energy transport mechanism and where the stellar pulsations originate. 
Dynamical wind models typically have an inner boundary situated just below the stellar photosphere, and therefore do not include any description of either convection or pulsation driving. 
However, the variation of the upper layers of the star is vital to wind driving as this induces the shock waves that facilitate dust formation. 
A parameterised prescription is typically used in the dynamical wind models, to simulate the temporal variation of the stellar luminosity and the radial velocity of the gas layers at the inner boundary. 
This prescription has historically had a simple sinusoidal form, based on the initial efforts to describe radial variation in AGB stars by \cite{bowen_dynamical_1988}, and has the advantages of few free parameters. 
However, self-excited interior pulsation models \citep[e.g.][]{ireland_dynamical_2011} and observations \citep[e.g.][]{nowotny_line_2010,lebzelter_shapes_2011} both indicate that this approach should be improved.

It has further been shown that assumptions made about the inner boundary conditions may have strong implications for the structure of the resulting model atmosphere, and consequently for the mass-loss rate and wind velocity, at least in the case of C-type AGB stars \citep{liljegren_dust-driven_2016}.

In this work we focus instead on self-consistent time-dependent models for dynamical atmospheres and dust-driven winds of M-type AGB stars.  
The models by \cite{hofner_winds_2008} and \cite{bladh_exploring_2013,bladh_exploring_2015} are based on a wind acceleration scenario by photon scattering on large, near-transparent silicate grains, which is supported by recent observational studies \citep[see e.g.][]{norris_close_2012,ohnaka_spatially_2012,ohnaka_clumpy_2016}.
Here we test a range of different inner boundary conditions for M-star atmospheric models. 
As the silicate dust present in the atmosphere of M-stars are more transparent than the amorphous carbon in C-stars, M-stars are generally less obscured by dust. 
There is therefore more observational material available for M-stars, e.g. high-resolution spectra, which facilitates comparisons between model results and observations.

To approach realistic pulsation properties and to pin down free parameters in the DARWIN models, we investigate the systematic effects of different pulsation properties (phase shifts between radial and luminosity variations, and different shapes of luminosity variations) on dynamical wind models. 
The resulting models are also compared to various observables. 
A set of 99 models with the same stellar parameters, but with a combination of different inner boundaries (see Sect. \ref{ib}), is calculated and then evaluated using three criteria:

\begin{itemize}
\item \textbf{Wind velocity and mass-loss rate -} A direct output from dynamical wind models (DARWIN models, Sect. \ref{darwin}), and a measure of the general dynamics. 
The models are evaluated by comparing the velocity and mass-loss rate combinations with observations from \cite{ olofsson_mass_2002} and \cite{gonzalez_delgado_``thermal_2003} in Sect. \ref{velcomp}. 
\item \textbf{Photometric variations -} The brightness in individual filters will vary during a pulsation cycle for AGB stars, due to various absorbers occurring throughout the atmosphere. 
This creates observable loops in colour-colour diagrams. 
The shape and positions of these loops will depend on the overall atmospheric structure and dynamics. Synthetic $(J - K)$ vs. $(V - K)$ loops are calculated for the DARWIN models, using the COMA code (Sect. \ref{coma1}), and are compared to observations in Sect. \ref{cc_comp}.
\item \textbf{High-resolution line profiles -} Vibration-rotation CO lines are formed in different layers of the atmospheres and wind acceleration regions. The Doppler-shifted CO second overtone line profiles and derived radial velocity (RV) curves reflect the velocity field of the pulsating photospheric layers of the star. Synthetic spectra for this line are calculated using the model atmospheres with COMA (Sect. \ref{coma2}) and resulting lines and RV curves are compared to observations in Sect. \ref{rvcomp}.
\end{itemize}

These tests are indicators of different aspects of the AGB atmosphere, and realistic models should reproduce the observations of all the three aspects at once. 
To the authors' knowledge this is the first study of its kind on M-type AGB star wind models.

The possibility of using the tests as diagnostics for pulsation models is also explored. 
The results from the  three tests are compared to pulsation properties derived from 1D self-excited radial pulsation models by \cite{bessell_phase_1996}, who provided a mathematical description in the form of Fourier series of the luminosity variation and the sub-photospheric dynamics from their models. 
This can be directly compared to the boundary condition used in the DARWIN models.
Furthermore, these pulsation models and follow-up simulations have been used extensively for the interpretation of various observations over the years, despite the lack of dust-driven winds in the models of \cite{bessell_phase_1996}. 

%
%

\section{Modelling methods}

\subsection{Dynamical wind models - DARWIN}
\label{darwin}

Pulsation, shock, and dust formation processes occurring in the atmospheres of AGB stars are inherently time dependent. 
To investigate the influence of different pulsation properties, atmosphere models are calculated using the Dynamical Atmosphere and Radiation-driven Wind models with Implicit Numerics (DARWIN) code. 
The DARWIN models simulate the time-dependent structures of an AGB star atmosphere using 1D frequency-dependent radiation-hydrodynamics, including a detailed treatment of dust growth and evaporation (for description see \citealt{hofner_dynamic_2016-1} and references therein).

For M-type AGB stars the winds are driven by photon scattering on large silicate (Mg$_2$SiO$_4$) dust grains, which grow through reactions with Mg, SiO and H$_2$O. 
The treatment of dust growth is time dependent, following the method described by \citet{gail_mineral_1999}, with the growth rate limited by the available Mg atoms. 
As the nucleation of dust grains in an oxygen-rich environment is not well understood \citep[see ][ and references therein]{gail_silicate_2016,gobrecht_dust_2016}, pre-existing seed particles consisting of 1000 monomers are introduced. 
The seed particle abundance is a free parameter and is defined as the ratio between number density of initial grains and number density hydrogen atoms. 
This is set to $3 \times 10^{-15}$ for all models calculated in this paper, following the findings in \cite{bladh_exploring_2015}.

The starting point for the simulations is a hydrostatic dust-free atmospheric structure whose fundamental parameters are effective temperature, luminosity, and chemical composition. 
The variations of the luminosity and the gas velocity at the inner boundary are gradually ramped up to simulate the pulsation of the star, transforming the hydrostatic initial atmosphere into a dynamical model. Using an adaptive spatial grid, the full system of non-linear partial differential equations describing gas, dust, and radiation is solved simultaneously with a Newton-Raphson scheme.

\subsubsection{Inner boundary conditions}
\label{ib}

   \begin{figure}
   \centering
   \includegraphics[width=\hsize]{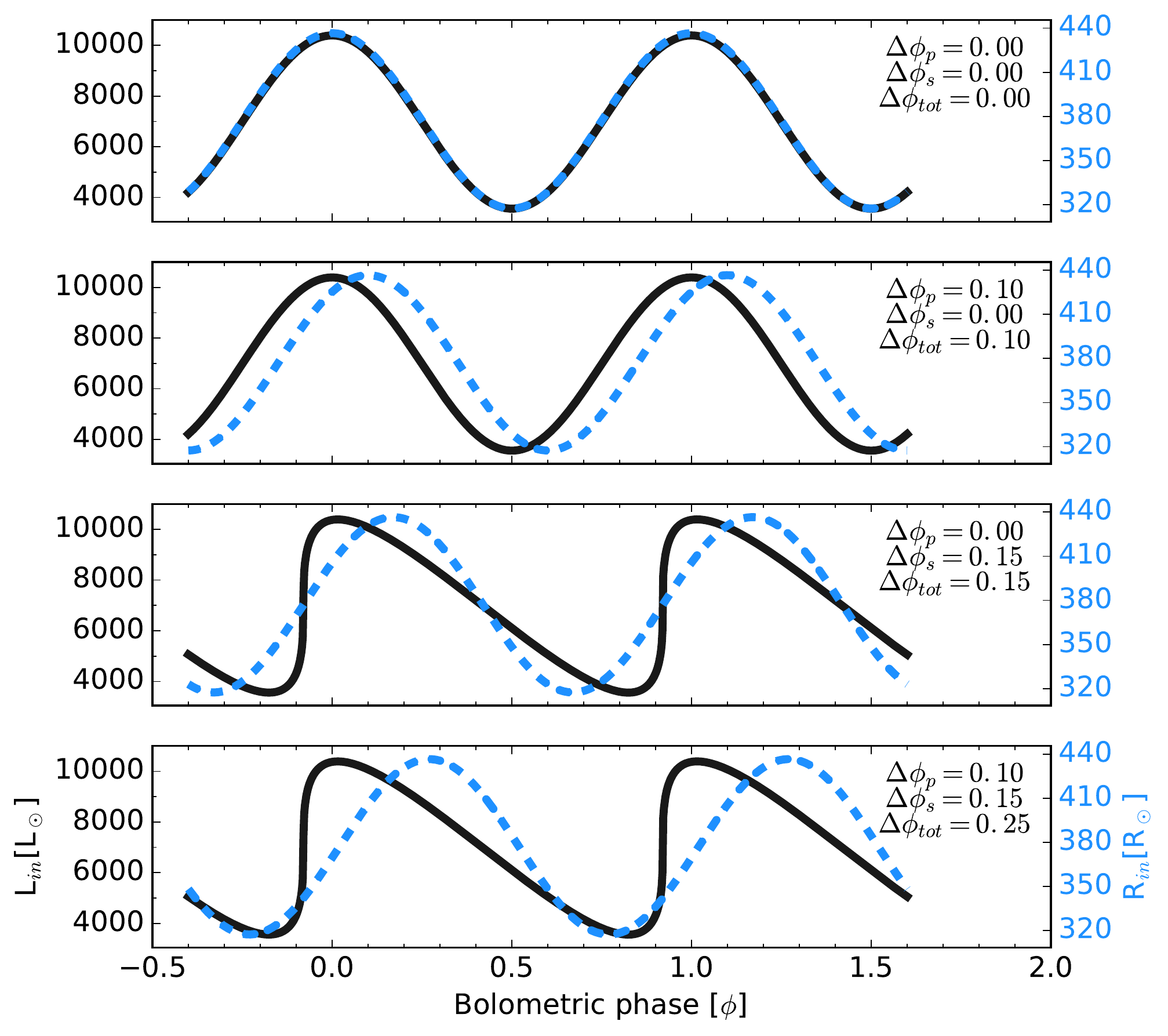}
      \caption{Different variations of the inner boundary condition used, showing luminosity variation (black) and the radial expansion and contraction (dotted blue). \textbf{Upper panel:} The original boundary condition, where the shape of the variation is sinusoidal for both luminosity and radial variation, and where the phase is locked. \textbf{Second panel:} Shape of the luminosity is unchanged, however an offset between the maximum luminosity and largest expansion is introduced, indicated by $\Delta \phi_p \neq 0$. \textbf{Third panel:} Shape of the luminosity is made asymmetric, indicated by $\Delta \phi_s \neq 0$.  \textbf{Lower panel:} Case with both offset ($\Delta \phi_p \neq 0$) and asymmetry ($\Delta \phi_s \neq 0$). }
         \label{bc_ex}
   \end{figure}

The spatial range of the DARWIN models, for models that develop a wind, reaches from just below the photosphere ($\sim 0.9R_\star$) out to where the wind has reached its terminal velocity $v_\infty$ ($\sim 20-30R_\star$). 
Since the inner boundary is located above the pulsation driving region, the variability of the stellar radius and the luminosity present in AGB stars is simulated with an ad hoc temporal variation of the physical quantities at the inner boundary. 
The form of these temporal variations has historically been very simple; $\Delta R_{in} \propto \sin(2 \pi t / P)$ for the radial variation and $\Delta L_{in} \propto \Delta R_{in}^2$ for the luminosity variation. 
Such a sinusoidal description, where the expansion and contraction of the stellar surface is locked in phase with luminosity variation, has been used in several different variations and generations of dynamical atmosphere and wind models \citep[e.g.][]{bowen_dynamical_1988,winters_systematic_2000,hofner_dynamic_2003,hofner_dynamic_2016-1}.

This simplified way of describing the time evolution of the luminosity at the inner boundary is not, however, representative of what is known about AGB stars pulsation properties. 
The shape of the luminosity variation is known to differ: \citet{lebzelter_shapes_2011} found that around 30$\%$ of Mira light curves deviated significantly from a sinusoidal curve and presented different shapes and secondary maxima. 
Interior pulsation models, by e.g. \cite{bessell_phase_1996} or \cite{ireland_dynamical_2008,ireland_dynamical_2011}, predict phase shifts between the variation of the surface layers and the luminosity. 
In \citet{nowotny_line_2010} atmosphere models of C-stars with the standard boundary condition were calculated; however, a synthetic phase shift had to be added to match the models to observations. 
To arrive at a better approximation of what a realistic boundary condition for the DARWIN models should be, we perform a systematic investigation of the influence of the inner boundary condition on various model properties, and compare the results with available observations.

The inner luminosity boundary is modified in two ways: i) by introducing a phase offset $\Delta \phi_p$ between the radial variation and luminosity variation and ii) by changing the shape of the luminosity variation, making it increasingly asymmetric (measured by $\Delta \phi_s$), as seen in panels 2 and 3 in Fig.\,\ref{bc_ex}. 
The result of both these changes is a phase shift between the maximum expansion and maximum luminosity of the star. 
The notation to describe these phase shifts introduced in \cite{liljegren_dust-driven_2016} is used here: $\Delta \phi_p$ for case i) (panel 2 in Fig.\,\ref{bc_ex}) and $\Delta \phi_s$ for the asymmetric case ii) (panel 3 in Fig.\,\ref{bc_ex}). 
The parameter $\Delta \phi_s$ is analogous to $\Delta \phi_p$, and measures the difference between the maximum stellar expansion and the maximum of the luminosity, which in case ii) is shifted because the luminosity variation is asymmetric. 
A larger $\Delta \phi_s$ then indicates an increasingly asymmetric light variation. 
The combination of these two changes will lead to a total phase shift between the maximum expansion and maximum luminosity as $\Delta \phi_{tot} = \Delta \phi_p+\Delta \phi_s$ (see panel 4 in Fig.\,\ref{bc_ex}).

For a full mathematical description of the boundary condition, see Appendix\,\ref{app1}. 

%
%

\subsubsection{Model set}
\label{mod_grid}

\begin{table}[t]
\caption{Parameters of the dynamical model \citep[adopted from][]{nowotny_line_2010} used in this study. } 
\label{tab1} 
\centering 
\begin{tabular}{l|cc}
\hline
Model: &  M2 \\ \hline
$L_{\star}$ {[}L$_{\odot}${]} & 7000 \\
$M_\star$ {[}M$_{\odot}${]} & 1.5 \\
$T_{\star}$ {[}K{]}  & 2600 \\
$R_{\star}$ {[}R$_\odot${]}  & 412 \\ \hline
{[Fe/H]} & 0 \\
C/O {[}by number{]} & 0.48\\ \hline
Period {[}days{]} &   490 \\
$\Delta u_p$ {[}kms$^{-1}${]}   & 6 \\
$f_L$  & 1.5 \\ \hline
\end{tabular}
\tablefoot{Model M2 simulates a typical M-type Mira developing a wind, with stellar parameter identical to those of \cite{nowotny_line_2010}, the only exception being the C/O abundance. Note that the stellar radius is calculated from luminosity and temperature. }
\end{table}

   \begin{figure}
   \centering
   \includegraphics[width=\hsize]{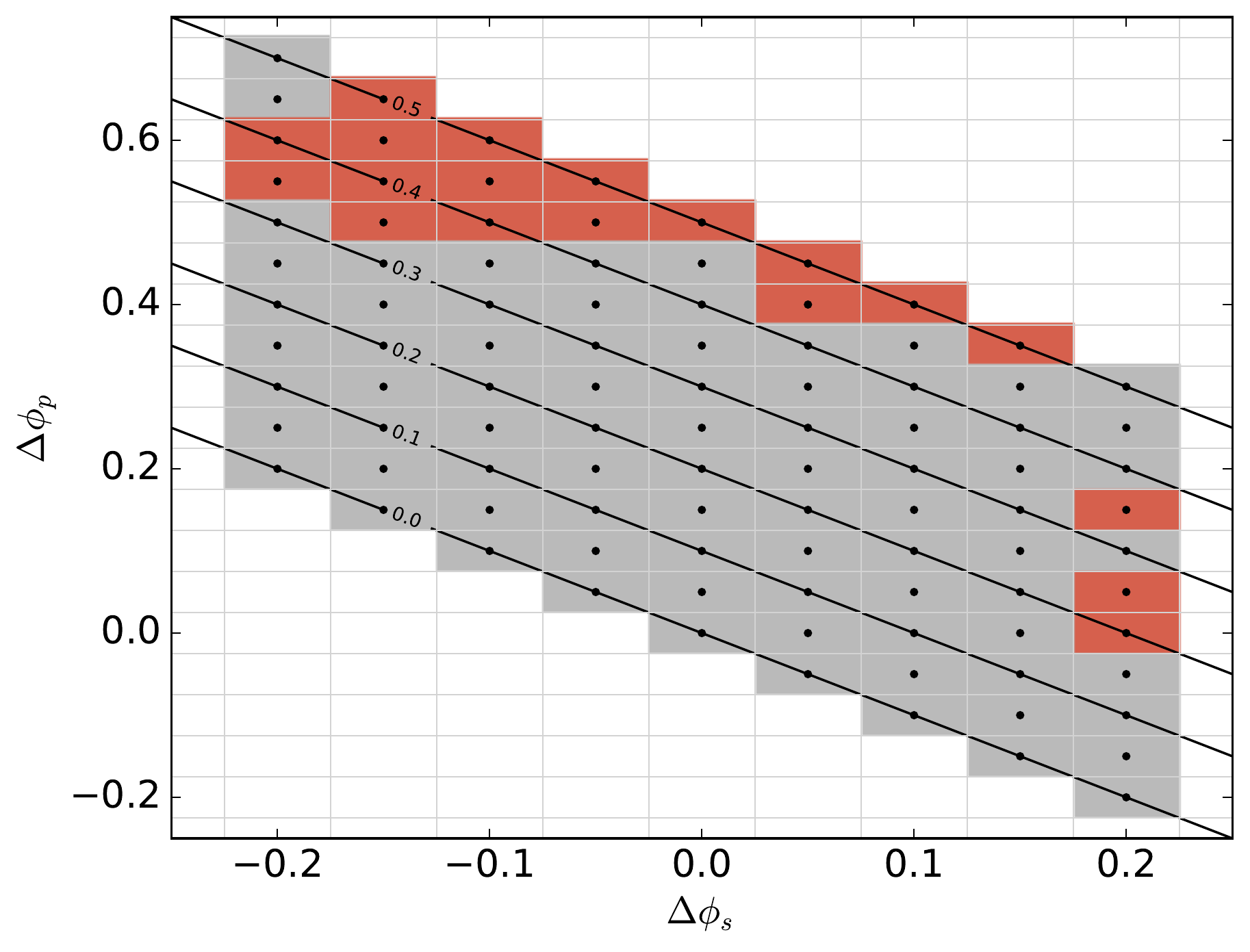}
      \caption{Overview of the calculated models, showing the $\Delta \phi_s$ and $\Delta \phi_p$ used. The points represent each of the models; grey background means the model finished while the red background means the model did not converge (Sect. \ref{mod_grid}). The lines show $\Delta \phi_{tot} = \Delta \phi_s+\Delta \phi_p $. }
         \label{mod_overview}
   \end{figure}

The model parameters used for the starting model (see Table \ref{tab1}) are based on stellar parameters of Model M in \citet{nowotny_line_2010}; the only difference is in the chemical composition (in previous work C/O$ = 1.4$, here C/O$= 0.48$). The atmospheric model here therefore has an O-rich chemistry, in contrast to the C-rich chemistry in the atmospheric model used in \citet{nowotny_line_2010}.
Continuing the naming tradition we call our model M2. 
This combination of parameters was found to reproduce very well the discontinuous S-shaped radial velocity (RV) curve, characteristic of Mira stars. 
The RV curve is derived from high-resolution time series spectra containing lines formed in regions with strong shocks (around $R \approx 1R_{\star}$), and can be used to deduce the dynamics of the inner layers of Mira atmospheres \citep[e.g. Sect.\,5.2 of ][]{nowotny_line_2010}.

\citet{leb02} found that RV curves of a sample of Miras, covering a range of periods between 145 and 470 days, showed a universal behaviour with very similar velocity amplitudes (difference between minimum and maximum velocities) and discontinuities in the RV curve around maximum bolometric phases. 
They also showed that while properties such as mass-loss rate and wind velocity differ between different Mira stars, the atmospheric dynamics of the inner layers seem to be similar, independent of metallicity, spectral type, period and chemistry.
It is therefore probably a good assumption that the result for model M2 can be generalised to other combinations of stellar parameters for Mira stars.

A set of 99 models with different inner boundary conditions for Model M2 is investigated. 
An overview of the set of models can be seen in Fig.\,\ref{mod_overview}. 
As observations \citep[see e.g.][]{nowotny_line_2010} and models \citep[see e.g.][]{ireland_dynamical_2011} both predict a positive phase shift, i.e. the stellar surface variation lags the luminosity variation, we constrain the investigation to cases where $\Delta \phi_{tot} \in [0, 0.5]$.

Of the 99 models, around 20$\%$ did not converge, in most cases due to the time step becoming very small. 
Such crashes are caused by extreme conditions (strong temporal changes) in models where the radial variation and luminosity variation are very much out of phase. 
Fortunately, the subsequent analysis showed that these models are far from the realistic region in parameter space, and will therefore not be discussed further.

\subsection{Spectral synthesis}
Photometry and high-resolution spectra are calculated for comparison with observational data. 
These calculations are done by applying an a posteriori approach, using resulting atmospheric structures from the DARWIN runs. 
For every model twenty snapshots per pulsation cycle, equidistant in time, are chosen. 
For each of these snapshots opacities are calculated using the COMA code, with the assumptions of LTE and a microturbulence of $\xi_t = $2.5 kms$^{-1}$.
Radiative transfer is then solved for the derived opacities. 
For a detailed description of the COMA code see \cite{aringer_synthetic_2016} and for a description of how the line profiles are calculated see \citet{nowotny_atmospheric_2005-1, nowotny_atmospheric_2005, nowotny_line_2010}.

%
%

\subsubsection{Photometry}
\label{coma1}

Comparisons using photometric colours are of interest as they trace variations of spectral energy distributions during a pulsation cycle. 
Large variations in the luminosity create strong variations in gas temperature, which in turn influence the molecular abundances in the atmosphere. 
These variations are especially prominent in the V-band, which is dominated by TiO, while H$_2$O affects the near-infrared wavelength region. 
Here the approach of \citet{bladh_exploring_2013,bladh_exploring_2015} is used, investigating the colour-colour diagram $(J-K)$ vs. $(V-K)$. 
Typically, observed M-stars show large variations in $(V-K)$ during a cycle, caused by changes in the molecule abundances (mainly TiO), while the variations in $(J-K)$ are small.

Observed photometric variations are represented by sine fits to light curves for the M-type Mira stars R Car, R Hya, R Oct, R Vir, RR Sco, T Col and T Hor \citep[for more details see][]{bladh_exploring_2013}. 
The difference between the observed loops are most likely due to differences in the fundamental properties of the stars, which is further explored in \cite{bladh_exploring_2013}, \cite{bladh_exploring_2015}, and \cite{hofner_dynamic_2016-1}.

Spectra for selected snapshots of every model are calculated, using a resolution of R $=10,000$ and covering a wavelength region between $0.3$\micron and $25$\micron.
We follow the same method here as described in \cite{nowotny_synthetic_2011} and \cite{aringer_synthetic_2016}.
Photometric filter magnitudes are calculated from these spectra, following the Bessell system \citep{bessell_jhklm_1988, bessell_ubvri_1990}. 
To be consistent with the observations, sine curves are fitted to the synthetic photometry variations of the models to produce loops in the $(J-K)$ vs. $(V-K)$ plane.

The observed loops can be compared to loops calculated from the models, as the dynamical model should be able to reproduce the general characteristics of these variations. 

%
%

\subsubsection{Line profiles in high-resolution spectra}
\label{coma2}

\begin{figure}
\centering
\includegraphics[width=\hsize]{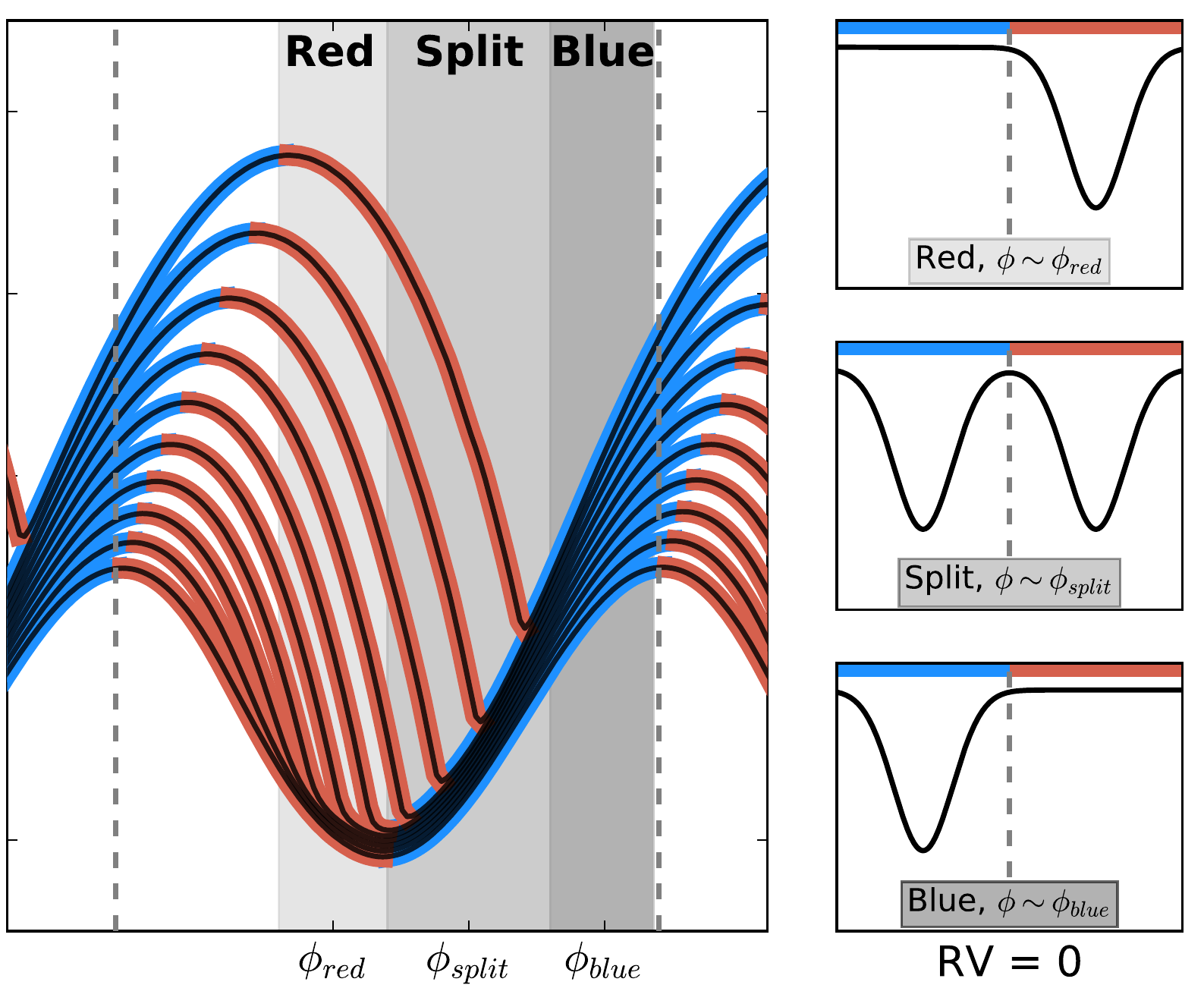}
\caption{Schematic of pulsating atmospheric layers, and lines created in this region at different times. The left panel shows mass-shells, and the colour indicates whether matter is infalling (red) or outflowing (blue).  
At times when the gas layers are moving inwards ($\phi \sim \phi_{red}$) the Doppler-shifted line will have a red component (first panel to the right), while outflowing material ($\phi \sim \phi_{blue}$) leads to a blue-shifted line (last panel to the right). When a shock waves moves through the line-forming region ($\phi \sim \phi_{split}$) there will be both a red component and a blue component (second panel to the right). }
\label{line_sh}
\end{figure}

The velocity field in the line formation region will affect the line profiles of molecular lines originating in the dynamical atmosphere. 
While the light curves computed in the previous section reflect the luminosity variations, the line profiles of molecular lines are affected by the velocity fields in the respective line-forming region. 
By calculating photometric variations and synthetic line profiles for the models and then comparing them with observations, it should be possible to deduce information about a potential phase shift between the luminosity and radial variation.

\citet{nowotny_atmospheric_2005-1,nowotny_atmospheric_2005, nowotny_line_2010} studied such line formation and velocity effects on various molecular features in C-type AGB star models. 
It was found that DARWIN models were able to reproduce various spectral features observed, and that there are three system of lines that are particularly interesting: the vibration-rotation CO lines $\Delta v = 1,2,3$.
These lines probe three different regions in the atmosphere as the temperature where these lines form are roughly 350-500K for CO $\Delta v = 1$, 800-1500K for CO $\Delta v = 2$, and 2200-3500K for CO $\Delta v = 3$. The CO $\Delta v = 1$ lines will then probe the outflow, while the CO $\Delta v = 2$ lines will probe the dust-forming regions and the CO $\Delta v = 3$ lines probe the dust-free layers \citep[e.g. Sect.\,5.1 of ][]{nowotny_line_2010}.

The CO $\Delta v = 3$ lines are formed in a region where the velocity field is dominated by effects of the shock waves, so the shape of the lines are determined by the propagation of the shock wave, which in turn depends primarily on radial variation due to pulsation. 
The movement of the mass-shells in such a line-forming region will then be imprinted on the line profile. A schematic of this scenario can be seen in Fig.\,\ref{line_sh}. 
When material is infalling the line profile will be red-shifted, seen at $\phi \sim \phi_{red}$  to the left in Fig.\,\ref{line_sh}.
As an outwards propagating shock wave hits the infalling material, there will be a line splitting due to the gas layers in the line-forming region moving both inwards (red-shifted component) and outwards (blue-shifted component).
This is seen at $\phi \sim \phi_{split}$. 
The line will be blue-shifted at $\phi \sim \phi_{blue}$ when all the layers are moving outwards.

Observations of Doppler-shifted lines in AGB stars will then directly probe the velocity field, and can be compared to model results. 
In \citet{liljegren_dust-driven_2016} such lines were synthesised for different variations of the inner boundary condition for the case of a C-star atmospheric model, and it was found that line profile variations, particularly the CO $\Delta v = 3$ line, combined with information about the visual phase, can be used as a diagnostic tool. This is tested here for the case of M-stars.

Line profiles for the CO line $\Delta v = 3$ (CO 5-2 P30 at 1.6573 \micron) at different times during a pulsation cycle are produced for DARWIN models with a range of boundary conditions. 
A resolution of R = 300,000 is used, covering a wavelength region between $1.6566$\micron and $1.6578$\micron. 
Frequency dependent opacities are produced using the COMA code, with assumptions that the line shapes are described by Doppler profiles. 
Gas velocities are taken into account when performing the radiative transfer because the velocity field, with both infalling and outflowing gas, defines the line profiles. 
Opacities due to dust sources are accounted for, however, they do not significantly contribute to the result as these dust species are very transparent. 
For a detailed description of the spectral synthesis see \cite{nowotny_line_2010}.

Synthetic RV curves are also calculated for each model, using the CO line $\Delta v = 3$ synthesis, and compared to observed radial velocity curves from \cite{leb02}. 

%
%

\subsection{Bolometric phase $\phi_{bol}$ and visual phase $\phi_{\rm v}$}

Two different measures of time are used throughout this paper: visual phase $\phi_{\rm v}$ and bolometric phase $\phi_{bol}$. 
Cycle dependent observations are usually linked to the star’s visual phase $\phi_{\rm v}$. 
For the model calculations we use the bolometric phase $\phi_{bol}$, which is always known, as one of the model outputs is the luminosity lightcurve.

For direct comparison between models and observations, when phases are important (Sect. \ref{rvcomp}), $\phi_{\rm v}$ is calculated for the models. 
For O-rich Miras the bolometric phase lags behind the visual phase $\phi_{\rm v}$, typically by $\phi_{bol}(max) - \phi_{\rm v}(max) \approx 0.1$ \citep[cf. Appendix\,A of ][]{nowotny_line_2010}. 
This is comparable to what is found in the simulations, where the differences between the visual and bolometric phase typically were in the range $0.05-0.15$ for the model atmospheres in this paper. 

%
%

\section{Resulting models}

\begin{figure*}
\centering
\includegraphics[width=\hsize]{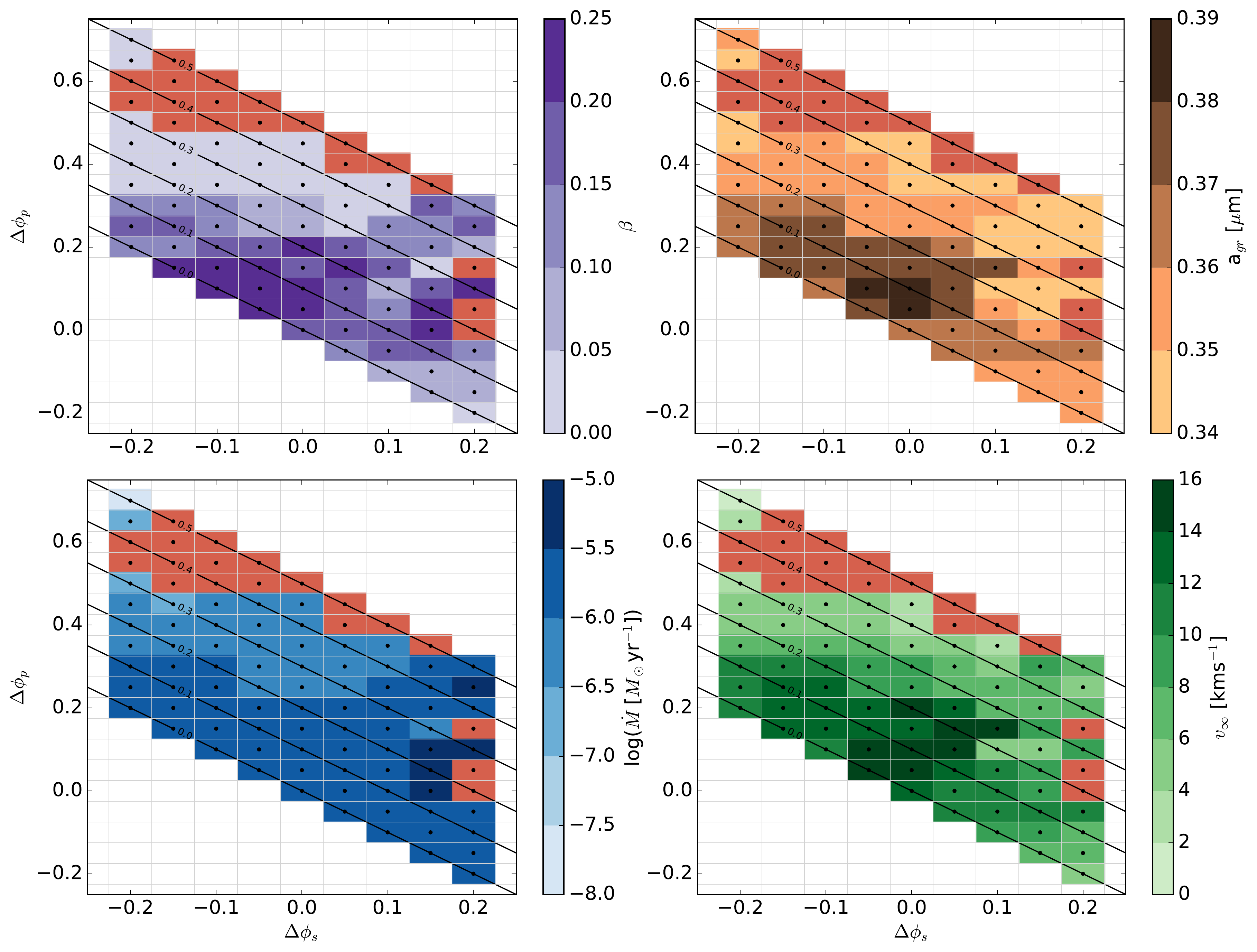}
\caption{Overview of the calculated models, showing the dependence on $\Delta \phi_s$, $\Delta \phi_p$ and $\Delta \phi_{tot}$. Each square represents one model, and the colours indicate physical properties. The red squares are models that did not converge (see Sect.\ref{mod_grid}). \textbf{Upper left:} $\beta$, the fraction of momentum available from the luminosity that has gone into driving the wind. \textbf{Upper right:} Final radius of the dust grains. \textbf{Lower left:} Log of the mass-loss rate. \textbf{Lower right:}Wind velocity of the models.}
\label{all_beta}
\end{figure*}

%
%

\subsection{Overview of the wind properties}

Changing the inner boundary may have large effects on efficiency of the wind driving in the resulting model atmosphere. 
With a mass-loss rate $\dot{M}$, which is due to the radiation pressure, the mass $\dot{M} \delta t$ ejected over a time interval $\delta t$ will acquire the momentum $\dot{M} \delta t v_\infty$ for a wind velocity 
$v_\infty$, by absorbing a fraction $\beta$ of the momentum available from the luminosity $L_\star \delta t /c $. 
Therefore $\dot{M} \delta t  v_\infty = \beta L_\star \delta t /c $, which can be solved for $\beta$ as

\begin{equation}
\label{eqn_beta}
\beta = \frac{\dot{M} v_\infty c }{L_\star}
\end{equation}

Here $\beta$ describes the fraction of momentum of stellar radiation that goes into driving the wind, and can be calculated from the outputs of the DARWIN models. 
The upper left panel of Fig.\,\ref{all_beta} shows $\beta$ for all models. 
Different boundary conditions result in a wide range of values of $\beta$, indicating that the form of the inner boundary is directly correlated with the efficiency of the wind driving in these models. 
There seem to be systematic effects; a group of models with $\Delta \phi _s \in [-0.15, 0.2]$ and $\Delta \phi _p \in [0, 0.2]$ are very efficient in wind driving. 
A large portion of the models with $\Delta \phi _p > 0.3$, on the other hand, are inefficient in driving the wind and produce much lower values of $\beta$.

The global properties of mass-loss rate and wind velocity are also highly dependent on the choice of the inner boundary conditions. 
The lower left and lower right panels of Fig.\,\ref{all_beta} show mass-loss rate and wind velocity plotted separately. 
Again some systematic effects can be seen; the models with $\Delta \phi _s \in [-0.1, 0.1]$ and $\Delta \phi _p \in [0, 0.2]$ result in the highest wind velocities, with values around $ v_\infty  = 12-16$ kms$^{-1}$. 
Velocities decrease with higher $\Delta \phi_{tot}$, with the lowest velocities being $ v_\infty  \approx 2$ kms$^{-1}$. 
There are also significant effects of changing the shape of the luminosity variation, specified by $\Delta \phi_{s}$.

Many of the models have a mass-loss rate between $ 1- 3 \times10^{-6}$ $M_\odot$ yr$^{-1}$, and do not seem to follow the same trends as wind velocity and $\beta$. 
There are several models with $\Delta \phi_s \in [0.1, 0.2]$ that have a relatively high mass-loss rate, but do not reach very high wind velocities. 
While a significant amount of material is lifted during the wind driving process for these models, the acceleration of this material is not efficient. 
The resulting wind velocities are too small to be realistic, in combination with the mass-loss rates (shown to the left in Fig.\,\ref{vel_dmdt_sig}). 
The lowest mass-loss rate is $\sim 10^{-8}$ $M_\odot$ yr$^{-1}$, which is two orders of magnitude smaller than the model with highest mass-loss rate.

The upper right panel of Fig.\,\ref{all_beta}, shows the radius of the silicate dust grains for all models, which is consistently between 0.35 and 0.4 \micron. 
As the dust radius $a_{gr}$ is related to the condensation degree of Si (denoted by $f_{cond}$) as  $f_{cond} \propto a_{gr} ^3$, this relatively small spread in grain radii still represents a significant difference in the degree of condensation.
The models with the smallest grain radii have condensation degrees of $f_{cond} \sim 0.15$, while the models with the largest grains have a  $f_{cond} \sim 0.3$
It is the models with the largest grain radii that also reach the highest wind velocities. 
This sets them apart from the group of models with high mass-loss rate but relatively low wind velocities ($\Delta \phi_s \in [0.1, 0.2]$), for which the grain radii are comparably small. 

%
%

\subsection{Atmospheric structure and dynamics due to different boundary conditions}

\begin{figure*}
\centering
\includegraphics[width=\hsize]{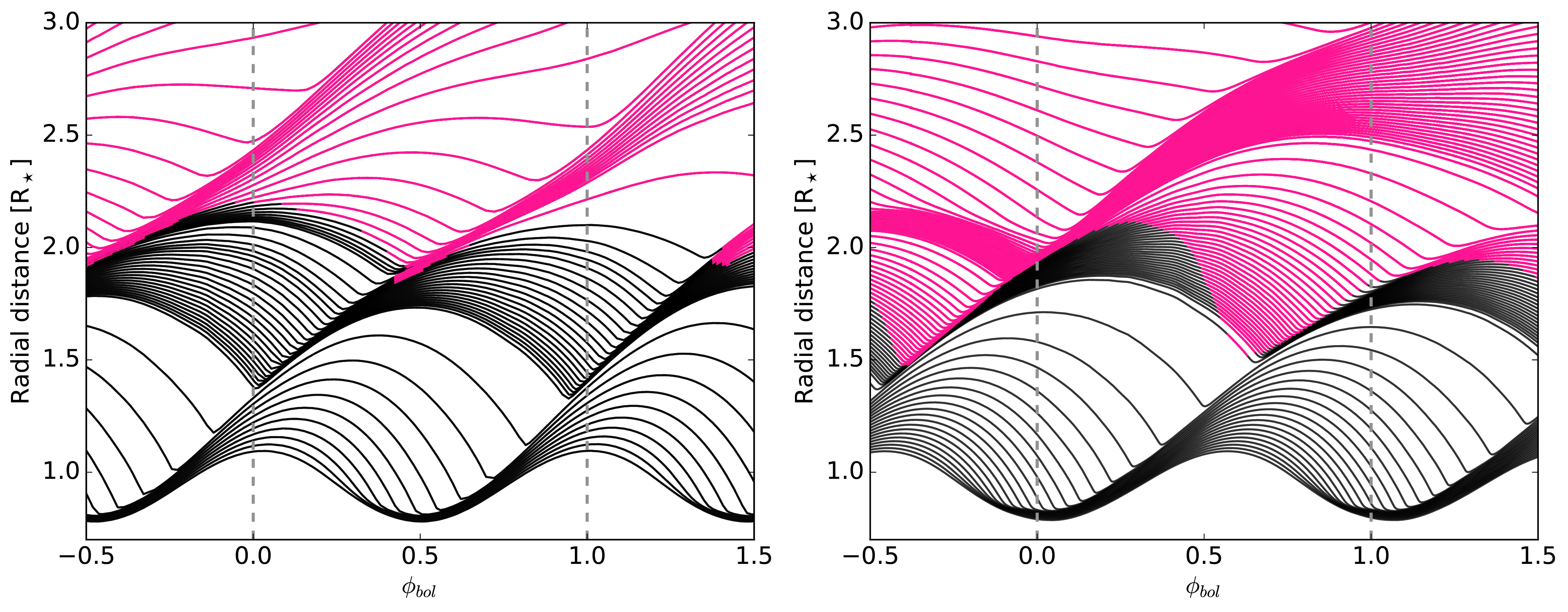}
\caption{Mass shell plots for two models. Pink represents where more than $1\%$ of the available Si has condensed, and indicates the presence of dust. \textbf{Left:} Standard model $\Delta \phi _p = 0.0$, $\Delta \phi _s = 0.0$; dust forms just behind the shock wave, where the density is high, leading to efficient wind driving. \textbf{Right:} Model $\Delta \phi _p = 0.4$, $\Delta \phi _s = 0.0$, with poor wind driving. Dust forms when matter is falling back onto the star, in a low-density region, resulting in significantly less matter condensing into dust, and therefore less material is accelerated in the wind. }
\label{ori_bad}
\end{figure*}

The large variety in mass-loss rates and wind velocities are due to the effect the inner boundary has on the structure of the atmosphere. 
This is discussed in detail for carbon stars in \citet{liljegren_dust-driven_2016}, and while the wind-driving dust species in C-stars is different from that of M-stars the mechanisms at play, determining the atmospheric structures and the wind properties, are comparable. 
A shorter explanation is given here by comparing a model with efficient wind driving (left panel in Fig.\,\ref{ori_bad}) to one with poor wind driving (right panel of Fig.\,\ref{ori_bad}).

The left plot in Fig.\,\ref{ori_bad} shows the atmospheric dynamics of the original model, where each line tracks the movement of a Lagrangian mass shell with time at different depths. 
This plot is illustrative of the processes contributing to a pulsation-enhanced dust-driven wind in AGB stars, indicating three distinctly different regions in the dynamical atmosphere. 
The dust-free region close to the surface at $R < 2 R_\star$ is dominated by the pulsation. 
When infalling matter collides with the outflowing gas, shock waves develop and propagate outwards. 
The material would follow ballistic trajectories if not for the onset of the wind by radiation pressure on the dust. Dust condenses at distances of $R \sim 2 R_\star$. 
In this dust-forming region the dynamical behaviour changes from strictly periodic to more sporadic, with some matter falling back onto the star and some material being accelerated outwards. A steady outflow is found beyond $R \sim 3R_\star$.

Ideally, for effective wind driving, the dust formation should take place in the wake of the propagating shock wave, where the dust can accelerate outward moving dense material further, leading to a strong wind. 
This is the case for the original model, seen in the left plot of Fig.\ref{ori_bad}, where dust is formed at $\phi_{bol} \sim 0.5$.
The dust formation region correlate with the lowest temperatures in the atmosphere, which in turn correlate with the luminosity minimum.

The most important change in atmospheric dynamics, when changing the inner boundary, is related to the phase of dust condensation. 
When shifting the luminosity variation compared to the radial movement of the upper stellar layers, which is the consequence of introducing new inner boundary conditions, the timing of dust condensation is also shifted. 
If the luminosity minimum and therefore the temperature minimum occurs such that dust is not formed when gas is moving outwards in a wake of a shock, but rather when material is falling back towards the star (seen in the right panel of Fig.\,\ref{ori_bad}), the wind driving becomes highly ineffective. 
This leads to significantly lower wind velocities and mass-loss rates.

Another important aspect, which also determines the effectiveness of the wind, is when the luminosity maximum occurs with respect to dust formation and shock wave propagation. 
A luminosity maximum earlier in the pulsation cycle leads to a larger radiative acceleration on the dust, and therefore a higher wind velocity and mass-loss rate. 
On the contrary, a luminosity maximum that is late with respect to the propagating shock wave means less acceleration and a lower wind velocity and mass-loss rate.

The combination of these two effects leads to the vast variety of different behaviours; there are diverse wind velocities and mass-loss rates for the different inner boundary conditions. This is the reason why some models are much more efficient in driving a wind. 

%
%

\section{Comparison to observables}

The models are evaluated via comparison to observations using three criteria: i) a combination of wind velocity and mass-loss rate, ii) colour-colour loops, and iii) RV curves derived from CO $\Delta v = 3$ vibration-rotation lines. 

%
%

\subsection{Wind velocity and mass-loss rate}
\label{velcomp}

\begin{figure*}
\centering
\includegraphics[width=\hsize]{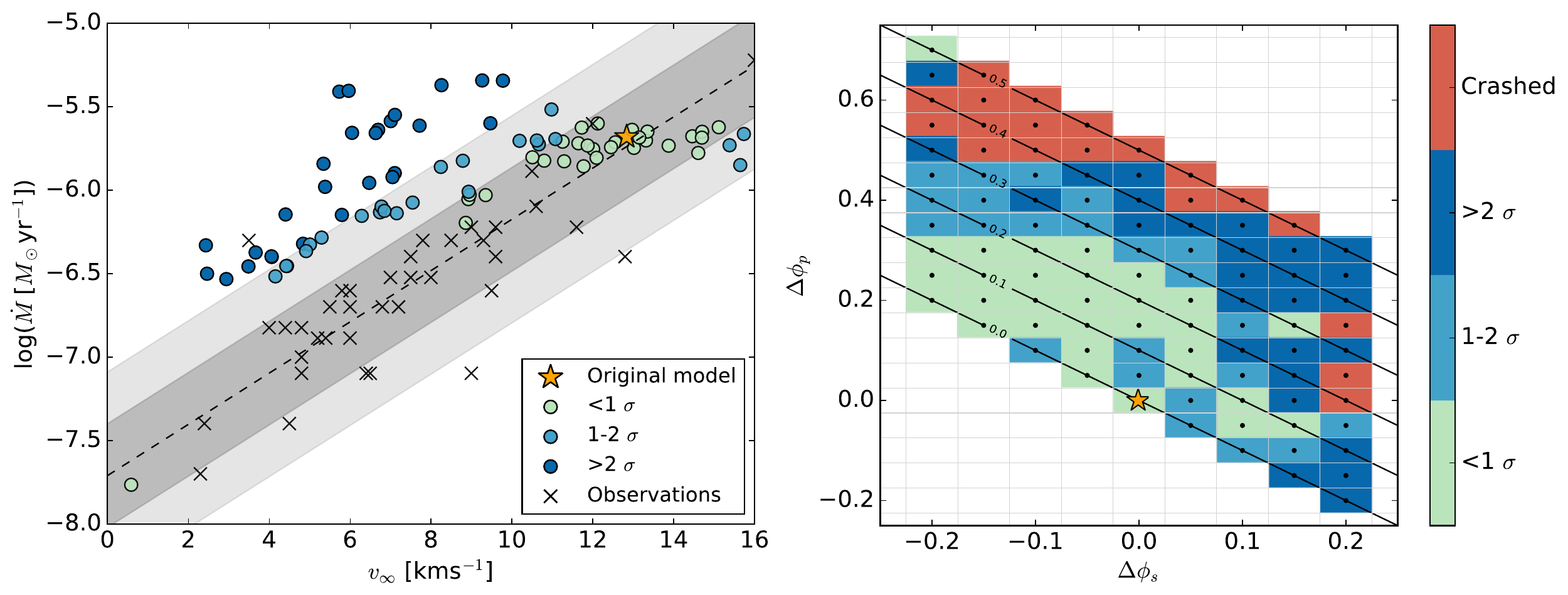}
\caption{\textbf{Left:} Mass-loss rates and wind velocities of models with phase shift at the inner boundary (circles), the original model (orange star) and observations (crosses). The dark grey area is the standard deviation $\sigma$ of the mass-loss rates of the observations, and the light grey is $2 \sigma$. \textbf{Right:} Each square represent a model; $\Delta \phi_s$, $\Delta \phi_p$ and $\Delta \phi_{tot}$ indicated. The colours are the same as in the left panel; models $<1 \sigma$ in the mass-loss rate and velocity plot are in light green. }
\label{vel_dmdt_sig}
\end{figure*}

Some of the most studied properties of AGB stars are the wind velocities and mass-loss rates, and DARWIN models should reproduce these properties realistically.

An overview of the wind velocities and mass-loss rates of the models is shown in Fig.\,\ref{vel_dmdt_sig} (pale green, blue, and dark blue circles and the original model as an orange star). This is compared to a linear fit to observational wind velocities against log of mass-loss rates for M-stars using observations from \cite{olofsson_mass_2002} and \cite{gonzalez_delgado_``thermal_2003} (crosses are observations, the dashed line is the fit to the observations). 
The observational uncertainties on the wind velocity vs. mass-loss rate relationship are dominated by uncertainties on the mass-loss rates, which are determined using CO multi-transitional line observations. 
The grey area plotted in Fig.\,\ref{vel_dmdt_sig} indicates the standard deviation from the linear fit in the mass-loss rates of the observations, which are comparable to the uncertainties found in \cite{ramstedt_reliability_2008} where the reliability of mass-loss rate estimates for AGB stars are investigated in detail. 
In the left panel of Fig.\,\ref{vel_dmdt_sig} the dark grey area marks the region with deviations from the linear fit being below 1$\sigma$, while the light grey area denotes a deviation between 1 and 2$\sigma$.

As previously mentioned changing the inner boundary may have large consequences for both the wind velocities and for mass-loss rates. 
The wind velocities range between approximately $1$ kms$^{-1}$ to almost $16$ kms$^{-1}$, while the mass loss can change over two two orders of magnitude, between $3 \times 10^{-6}$ and $10^{-8}$ $M_\odot$ yr$^{-1}$.

For higher wind velocities, the wind seems to saturate at some maximum mass-loss rate, in this case around $1.5 \times 10^{-6}$ $M_\odot$ yr$^{-1}$. This leads to a flat distribution in mass-loss rate above $10$ kms$^{-1}$. The changes to the inner boundary thus have little consequence for the mass-loss rate beyond this point; however, the wind velocity is still sensitive to such changes. 
Below $10$ kms$^{-1}$ there is a wider spread, but the models seem to have mass-loss rates that are too high for such low wind velocities, which is not realistic when compared to the observations.

The right panel in Fig.\,\ref{vel_dmdt_sig} shows the comparison between the observations and the models for different combinations of boundary conditions. As seen, there is a group of models with $\Delta \phi_{tot} \in [0.0, 0.3]$ that produces realistic combinations of wind properties. The models that did not converge (red) are relatively far from this group of models.

There are in total 30 models that simultaneously reproduce a realistic combination of mass-loss rates and wind velocities within $1\sigma$. 
Most of these models have combinations of $\Delta \phi _s \in [-0.2, 0.15]$ and $\Delta \phi _p \in [0, 0.3]$, with $\Delta \phi_{tot} \in [0,0.3]$. 
It should be noted that the model with $\Delta \phi _s = -0.2$ and $\Delta \phi _p = 0.7$, one of the most extreme models, has a very low mass-loss rate and a very low wind velocity($\dot{M} \approx 10^{-7.7}$$M_\odot$yr$^{-1}$, $v_\infty\approx 1$kms$^{-1}$), and does not reproduce Mira-like properties. 
This model is therefore excluded, even though it technically falls within the $1 \sigma$ range.

The 29 models left are used for further analysis and to investigate the colour-colour diagrams and high-resolution spectra. 
The model with the standard boundary conditions is among these 30 stars, and it produces a realistic mass-loss rate and wind velocity combination. 
This is rather reassuring, considering that the stellar parameters of this model emerged from a study based on C-rich dynamical models (with a different wind-driving species).
As mentioned in Sect. \ref{mod_grid}, the pulsation properties of stars that have different chemical compositions  but otherwise similar stellar properties should be comparable.  

%
%

\subsection{Loops in colour-colour diagrams}
\label{cc_comp}
   \begin{figure}
   \centering
   \includegraphics[width=\hsize]{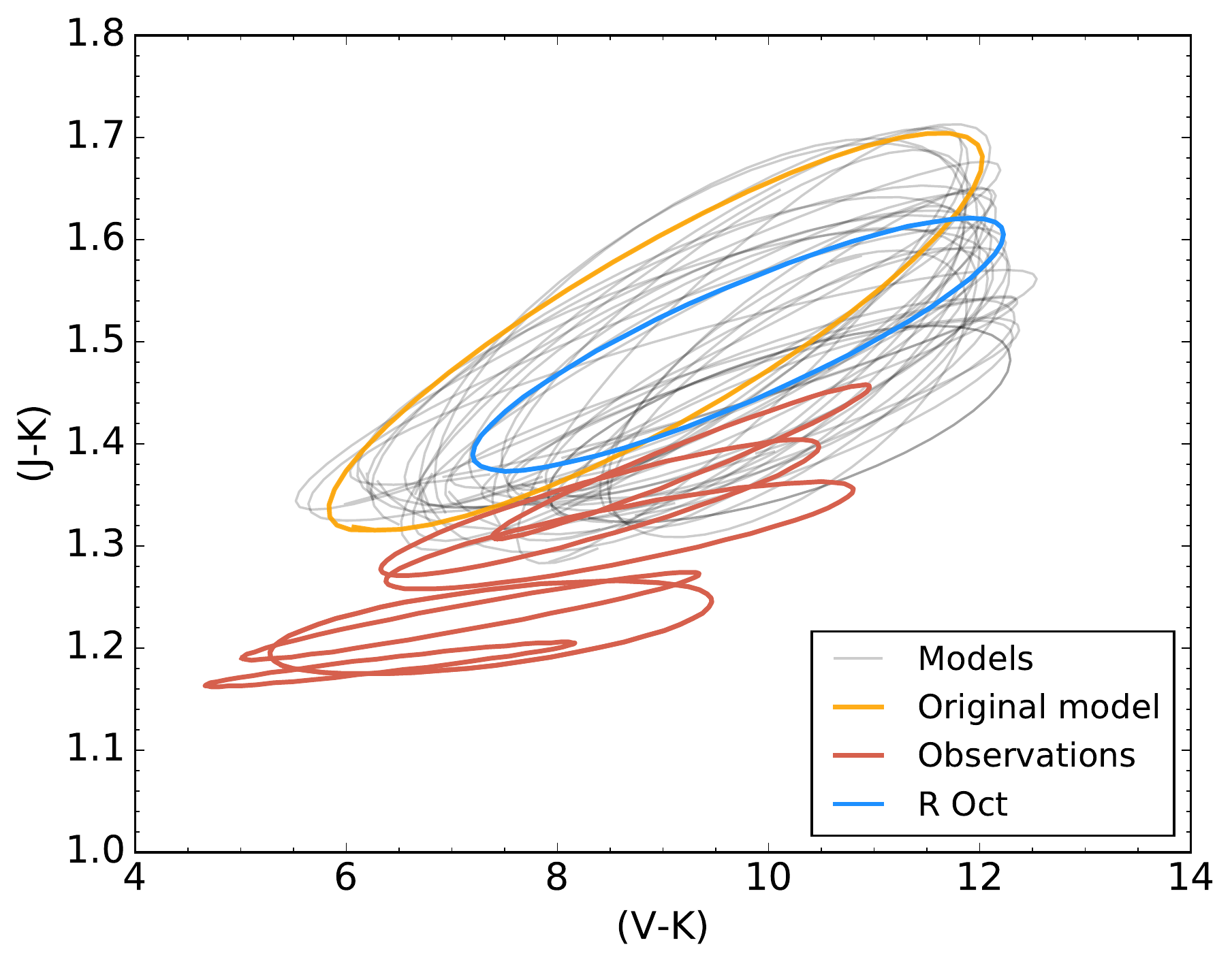}
      \caption{Colour-colour loops in the (J-K) vs (V-K) plane, with models in grey, the standard model with no phase shift in orange, and observations of seven Mira variables shown in red and blue \citep[see ][for more details]{bladh_exploring_2013}. R Oct (blue) seems to be very similar to the synthetic loops }
         \label{all_cc}
   \end{figure}

\begin{figure*}
\centering
\includegraphics[width=\hsize]{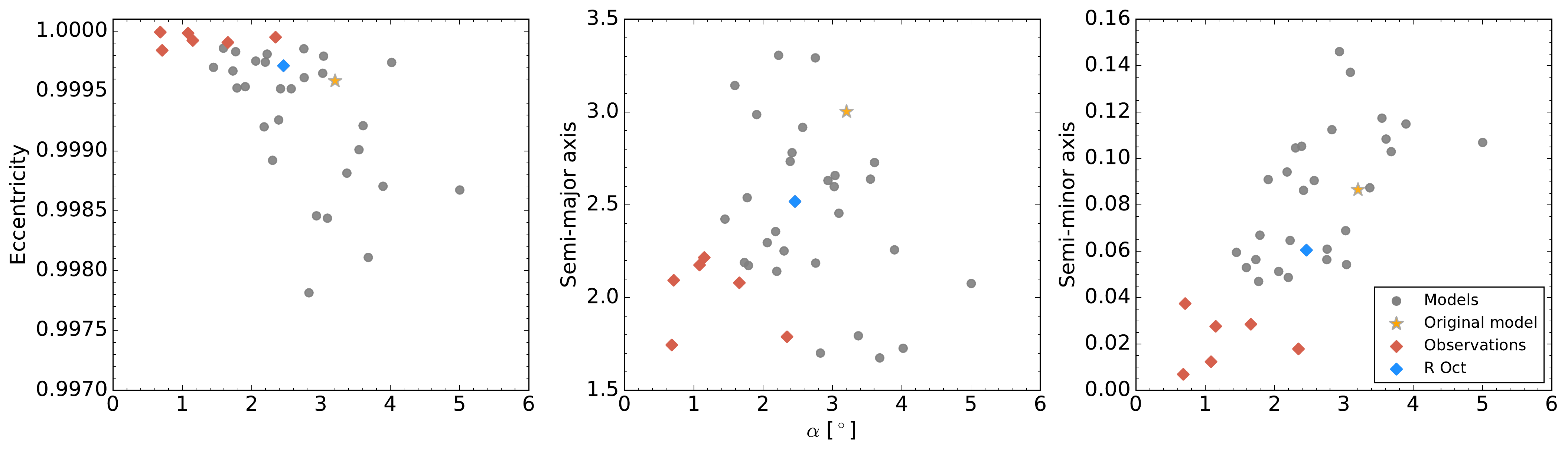}
\caption{Ellipse properties of the synthetic (grey) and observed (red and blue) colour-colour loops, in the $(V-K)$ vs $(J-K)$ plane against the angle of inclination. \textbf{Left:} Eccentricity, a measure of the shape. \textbf{Middle:} Semi-major axis, $(V-K)$. \textbf{Right:} Semi-minor axis, $(J-K)$. }
\label{ell_prop}
\end{figure*}

The photometric comparison with observations is done by calculating synthetic colours for the 29 models that produced satisfying mass-loss rates and wind velocities. 
Fig.\,\ref{all_cc} shows the loops formed in the $(J-K)$ vs. $(V-K)$ diagram after sine curves are fitted to the photometric variations, for both observed stars (red and blue loops) and for synthetic colours (grey loops). 
The model with the original boundary condition is shown in orange.

%

Models with higher $T_{\rm eff}$ and lower mass-loss rates, which have been studied in previous papers, fell mostly into the area covered by loops shown in red \citep[see e.g. Fig. 8 in ][]{hofner_dynamic_2016-1}. Model M2, with its lower $T_{\rm eff}$ and higher mass-loss rate, on the other
hand, is more comparable to R Oct (shown in blue), which has the highest mean ($J - K$) and ($V - K$) in the observed sample. The ($J - K$) values are commonly considered to be an indicator of $T_{\rm eff}$, which is also reflected by models 
\citep[see Fig. 7 in][]{bladh_exploring_2013}.   
The ($V-K$) values depend on the strength of molecular features (mostly TiO), which in turn are strongly affected by the variable structure of the dynamical atmosphere and the condensation distance of the wind-driving dust grains 
\citep[see the discussion in][]{bladh_exploring_2013}.   
Observations and models both loop in an anti-clockwise direction in this diagram, indicating qualitatively similar phase lags between light curves in the different photometric bands.


In an attempt to more quantitatively examine how well the observations and the model results agree and to see if any models significantly deviate, we describe the loops as ellipses in the $(J-K)$ vs. $(V-K)$ plane. 
An ellipse is defined by five quantities: the position of its centre $(x_0, y_0)$, the semi-major axis $a$, the semi-minor axis $b$ and the angle of inclination $\alpha$ . The shape of an ellipse is described by its eccentricity as 
$e = \sqrt{1 - b^2/a^2}$.


%
%

The centre positions of the loops of the 29 models, which were selected based on their wind properties, are within the range of observed ($J - K$) and ($V - K$) colours for larger observational samples \citep[mostly single epoch data, see][]{bladh_exploring_2013}, and can therefore not be used to put extra constraints on the boundary conditions. They will not be discussed further. 
Eccentricity, semi-major axis $a$, the semi-minor axis $b$ are plotted against the angle of inclination $\alpha$ for model loops and for observed loops in Fig.\,\ref{ell_prop}, again with observations in red and blue and models in grey with the original model as the orange star.

The left panel in Fig.\,\ref{ell_prop} shows the eccentricities for the observed loops, which are all very close to 1. 
This very high eccentricity is due to a small variation in the $(J-K)$ plane compared to the variation in the $(V-K)$ plane, and common for all the observed loops. 
The angle of inclination $\alpha$ is between  $0^\circ$ and $3^\circ$, meaning the observed loops are very narrow and almost flat. 
The model loops have a spread in both eccentricity and angle of inclination; however, some of the models overlap very well with the observations.

The middle and right panels of Fig.\,\ref{ell_prop} show the semi-major and semi-minor axis respectively. 
Again, there is a spread for the model loops, but it seems to be realistic as this variation is similar to the spread between the different observations. R Oct seems to be a good fit for both the semi-major and semi-minor axis, and the eccentricity. This indicates that models can reproduce realistic $(J-K)$ (semi-minor) and $(V-K)$ (semi-major) variations simultaneously.

There are no models that obviously deviate from the observations. The spread of the different properties for the models is comparable to the spread of the observations. 
Therefore, all 29 models are used in the further analysis of the line profile variation.

%
%

\subsection{Line profiles in high-resolution spectra}
\label{rvcomp}

High-resolution spectra were calculated for the sample of 29 models that produced realistic mass-loss rates and wind velocity combination as well as realistic colour variations.  
Synthetic RV-curves from the spectra are compared to observed RV curves for Mira AGB stars.

\begin{figure*}
\centering
   \includegraphics[width=\hsize]{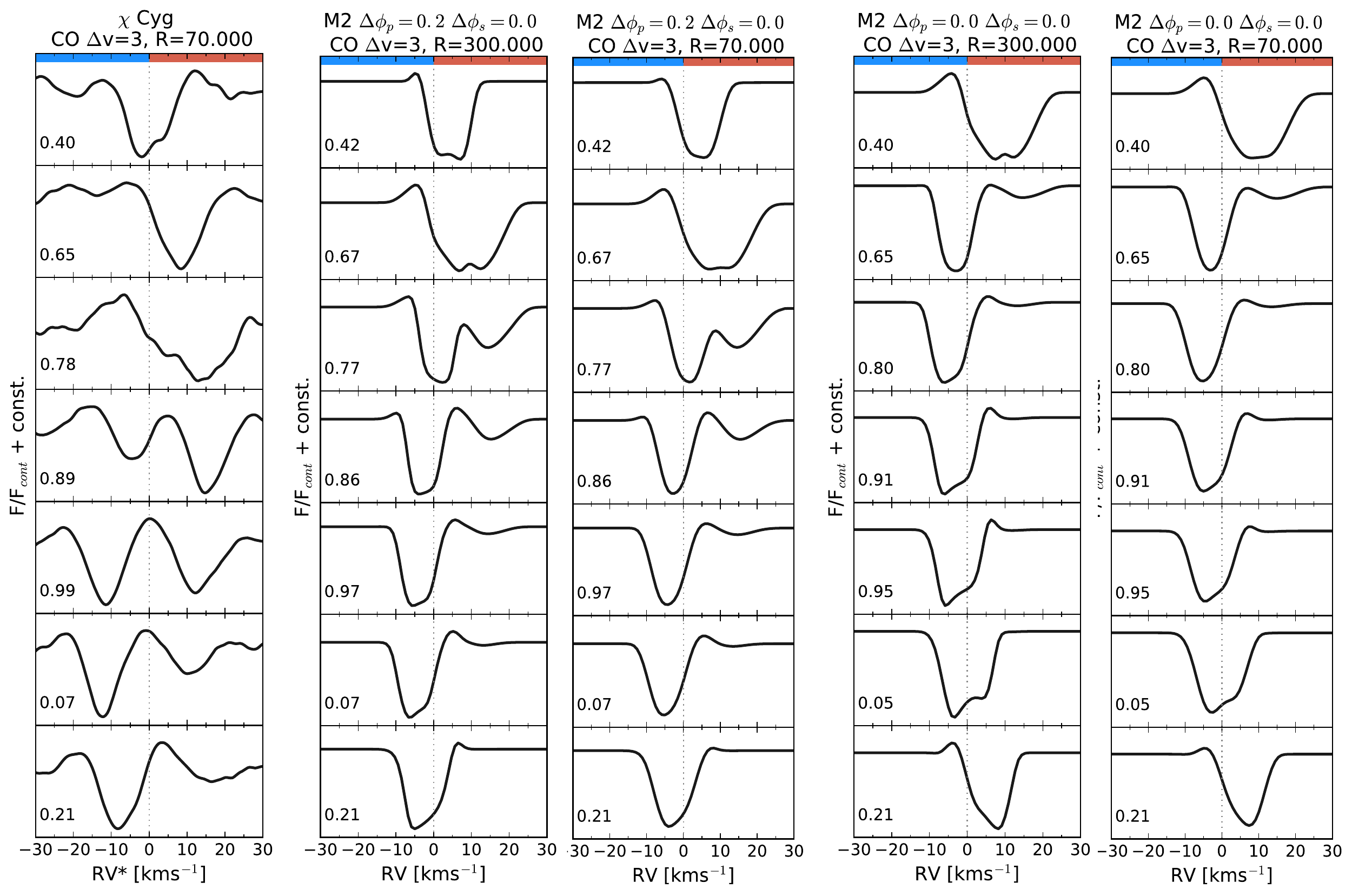}
      \caption{Line profiles from observations and from synthesis. The number is $\phi_{\rm v}$ of that panel. \textbf{Column 1 from the left:} Time series of an average of 10-20 unblended CO $\Delta v = 3$ lines, using FTS spectra of $\chi$ Cyg, taken from \citet{lebzelter_4_2001}. Observed heliocentric velocities were converted to systemic velocities RV* using heliocentric centre of mass radial velocity CMRV$ = -7.5kms^{-1}$ from \citet{hinkle_time_1982}. \textbf{Column 2 and 3:} Synthetic CO $\Delta v = 3$ (5-2 P30) line profiles, from a model with $\Delta \phi_{tot} = 0.2$, over a full cycle, shown for resolution 300,000 (Col. 2) and 70,000 (Col. 3).  \textbf{Column 4 and 5:} Synthetic CO $\Delta v = 3$ (5-2 P30) line profiles, from the original model, over a full cycle, shown for resolution 300,000 (Col. 4) and 70,000 (Col. 5).   }
         \label{lines3}
   \end{figure*}

\subsubsection{Investigating the inner atmosphere with CO $\Delta v = 3$}

The CO $\Delta v = 3$ lines originate deep in the atmosphere below the dust-forming layers where the radial expansion and compression of the upper stellar layers induce strong shock waves. 
The movement of the mass-shells in such a line-forming region will then be imprinted on the line profile, as described in Sect. \ref{coma2} and illustrated in Fig.\,\ref{line_sh}.

This behaviour is observed for CO $\Delta v = 3$ lines in AGB stars.
The first column of Fig.\,\ref{lines3} shows time series observation of CO second overtone lines (average of 10-20 unblended lines) of $\chi$ Cyg, a S-type Mira variable. 
The $\phi_{split}$ of $\chi$ Cyg (described in Fig.\,\ref{line_sh}) occurs at $\phi_{\rm v} \sim 0$. 
This line splitting at maximum visual light seems to be a ubiquitous feature of for Mira AGB stars.

Such line splitting and the resulting RV curve can also be synthesised using DARWIN models. 
The line profiles for the standard model, $\Delta \phi_p =0$, $\Delta \phi_s =0$, are shown in the two rightmost panels of Fig.\,\ref{lines3} at different resolutions. 
The shape of the line changes during one pulsation cycle; when material is outflowing the line has a strong blue-shifted component ($\phi_{\rm v} \approx 0.65 - 0.0$), while the infalling material results in a red-shifted component ($\phi_{\rm v} \approx 0.2-0.4$). The $\phi_{split}$ in such a model occurs at $\phi_{\rm v} \sim 0.8$, which is too early compared to observations.

There is very little difference between the models with different boundary conditions concerning the actual line shape during a cycle other than the timing of different features, such as the line-splitting, occurring at different $\phi_{\rm v}$. This can be seen by comparing panels four and five (the original model, with $\Delta \phi_{tot}= 0.0$) with panels two and three ($\Delta \phi_p = 0.2$, $\Delta \phi_s = 0.0$, with $\Delta \phi_{tot} =0.2$). The line profiles of the two models are similar, but different line features occurs at different $\phi_{\rm v}$.
This indicates that changing the luminosity boundary has few consequences for the shape of the line, which in turn demonstrates that the material at these radial distances is not subjected to significant radiation pressure.

The main characteristics of the observed line, with a line-splitting around $\phi_{\rm v} = 0.0$ and the overall shift of the main component of the line, can be reproduced if a model with $\Delta \phi_{tot} =0.2$ is used (such as shown in panels two and three).
The velocity field of this model is then in overall agreement with that of the observed star. 
There are some differences however; the red component during the line splitting is significantly weaker for the models, probably due to a lower density of the infalling material.

\begin{figure}
\centering
\includegraphics[width=\hsize]{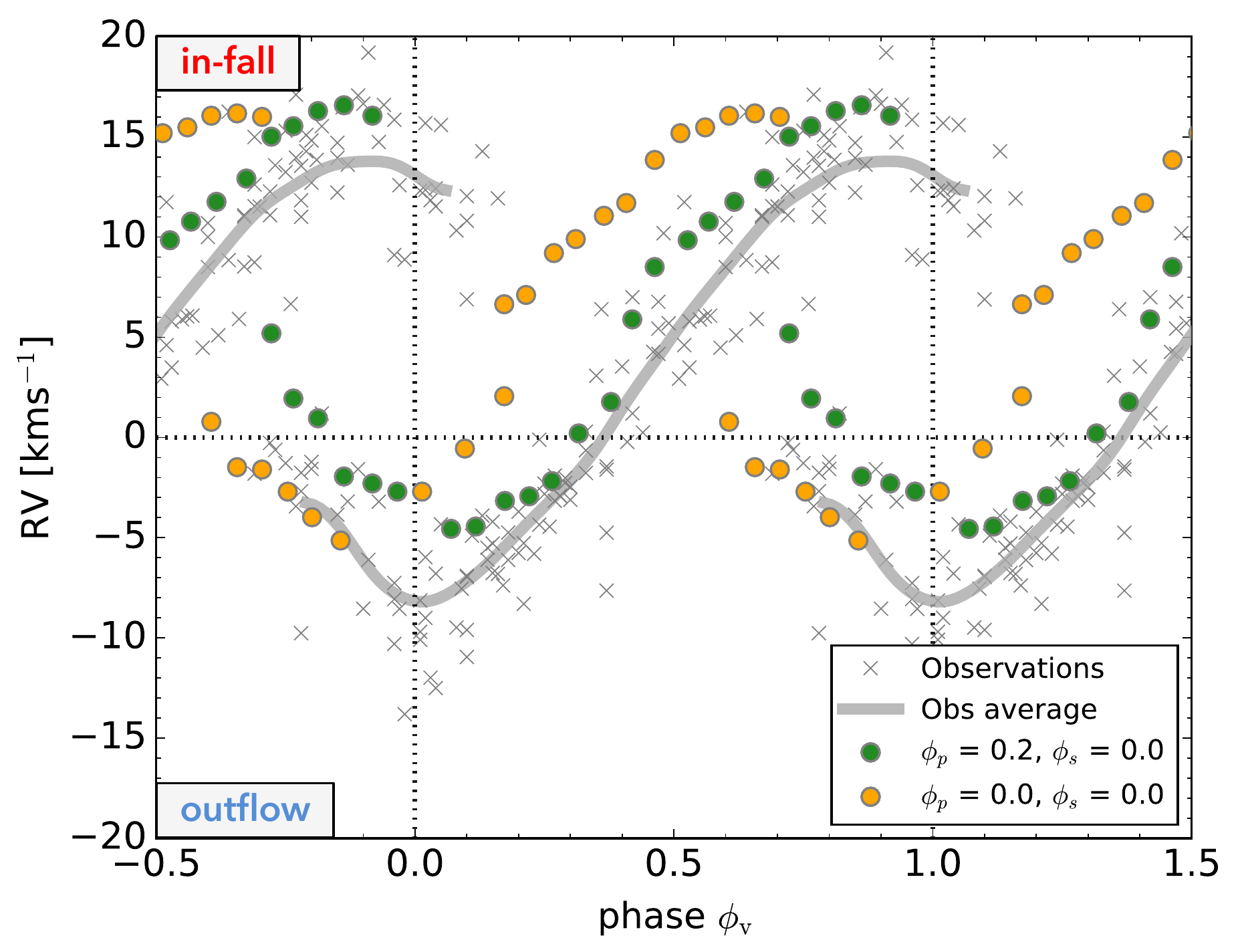}
\caption{Radial velocities (RVs), with circles showing RVs derived from synthetic lines for two models, observation as crosses, and the grey line a running mean for the observations. The observed measurements are a compilation of RVs derived from FTS spectra for a large sample of different Mira stars, with the RVs converted to systematic velocities using the centre of mass radial velocity (CMRV) for each object \citep[data from ][]{leb02}.}
\label{rv_overview}
\end{figure}

\begin{figure}
\centering
\includegraphics[width=\hsize]{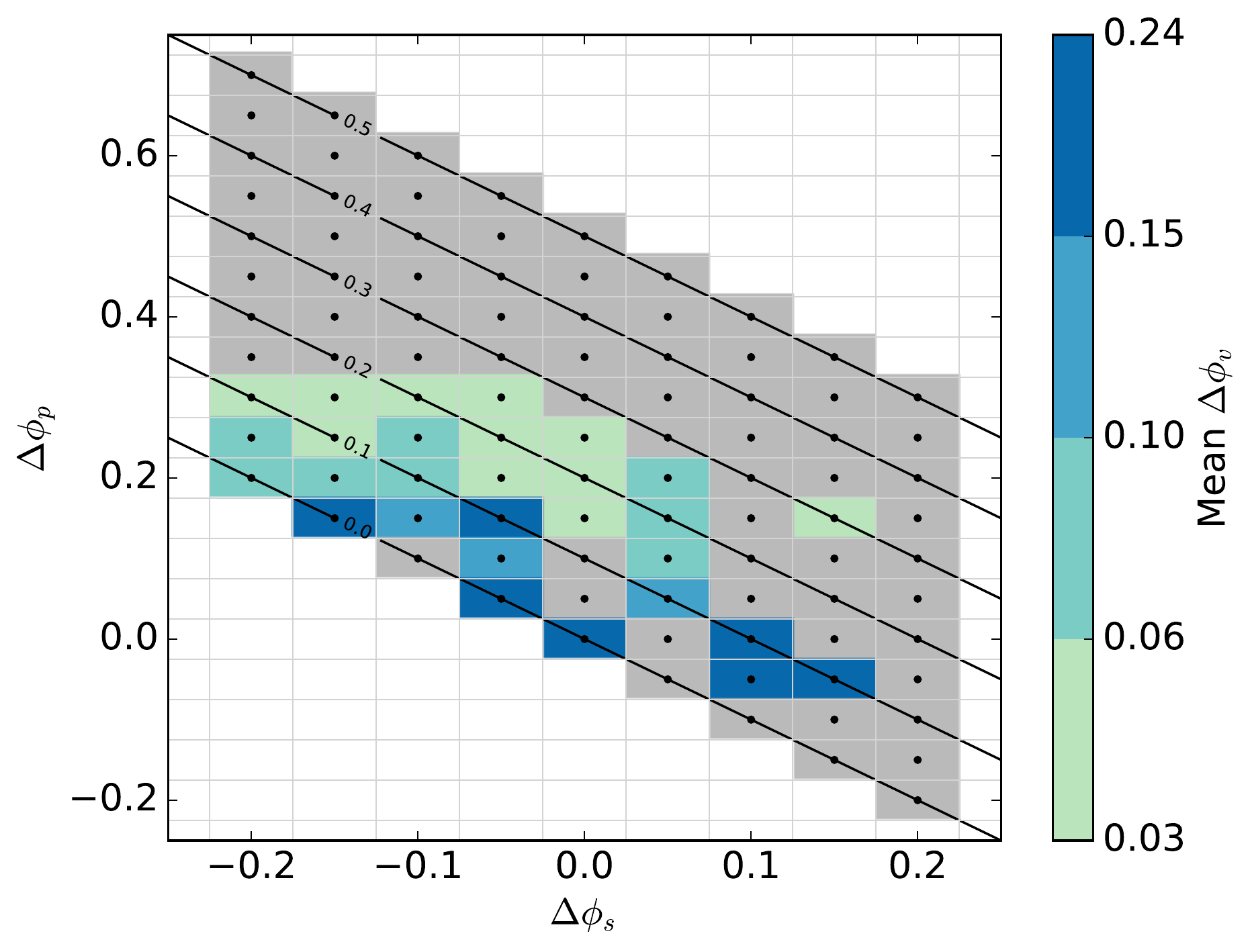}
\caption{Comparison between synthetic RV curves of models and the running mean of the observations, seen in Fig.\,\ref{rv_overview}. The colour indicates the mean distance between the maximum and the minimum of that model's RV curve and the running mean, described in Eq. \ref{diff}. The grey squares are models that were previously discarded as they did not reproduce a realistic wind velocity and mass-loss rate combination. }
\label{rv_comp}
\end{figure}

\subsubsection{Comparison of radial velocity curves}
\label{srv_comp}

Radial velocity curves derived from observations of second overtone CO lines in Mira stars show a uniform behaviour: a discontinuous S-shaped curve, with both outflowing and infalling components at $\phi_{\rm v} = 0.0$, $RV=0$ at around $\phi_{\rm v}=0.4$ and a velocity amplitude $\Delta RV \approx 25$ kms$^{-1}$ (Fig.\,\ref{rv_overview}, crosses and grey fit, which is the running mean over $\phi_{\rm v}= 0.2$ for the observation).
The shape and amplitude of the observed RV curve are very well reproduced in previous works, deriving RV curves from synthetic line spectra of C-star models; however, both the line splitting and $RV=0$ of these models occur earlier than $\phi_{\rm v} = 0.0$ and $\phi_{\rm v}=0.4$ respectively \citep[see e.g. ][]{nowotny_atmospheric_2005,nowotny_line_2010}. 
The same problem is found for the standard M-type model here, when calculating the RV curves from the CO $\Delta v  = 3$ line synthesis, indicated by orange dots in Fig.\,\ref{rv_overview}.

Models with a larger $\Delta \phi_{tot}$ approach observations: a line-splitting occurs closer to $\phi_{\rm v} = 0.0$. 
This can be seen in Fig.\,\ref{rv_overview}, where the green dots represent a model with a larger total phase shift, $\Delta \phi_{tot}=0.2$ ($\Delta \phi_p =0.2$, $\Delta \phi_s =0.0$). 
Increasing $\Delta \phi_{tot}$ of the model will essentially shift the RV curve to the right. 
So while the inner luminosity boundary does not affect the line-forming regions directly, it does change the visual phase, which is important when comparing models to observations.

We want to compare the resulting RV curves for the 29 models that have realistic wind velocities and mass-loss rates with the mean RV curve from the CO-line observations of Mira stars. 
In this case it is not suitable to use a standard L2 norm (a measure of the mean quadratic difference between the points on two curves, for mathematical description see Appendix \ref{a3})  as an error estimation.
Results from such an error might be misleading, due to the offset between the model curves and observational curve and due to the discontinuous nature of the RV curve. 
Instead we compare the timing of the synthetic and observed RV curves' minimum ($\phi_{\rm v}(min)$),  maximum ($\phi_{\rm v}(max)$), and the RV zero point ($\phi_{\rm v}(RV=0)$). 
These points are defined as $\phi_{\rm v}(s, min)$, $\phi_{\rm v}(s, max)$, and $\phi_{\rm v}(s, RV=0)$ for the synthetic RV curves and $\phi_{\rm v}(o, min)$, $\phi_{\rm v}(o, max)$, and $\phi_{\rm v}(o, RV=0)$ for the observed mean curves.
The differences between the synthetic and observed RV curve for these points are defined as $\Delta max = |\phi_{\rm v}(o, max) - \phi_{\rm v}(s, max)|$, $\Delta min = |\phi_{\rm v}(o, min) - \phi_{\rm v}(s, min)|$ and $\Delta zero = |\phi_{\rm v}(o, RV=0) - \phi_{\rm v}(s,RV=0)|$.
We then define the mean difference as

\begin{equation}
\label{diff}
\Delta \phi_{\rm v} = \frac{\Delta max+\Delta min+\Delta zero}{3}.
\end{equation}

The value of $\Delta \phi_{\rm v} $ is calculated for the 29 models. 
The results are shown in Fig.\,\ref{rv_comp}. 
To have a line splitting occurring close to what is observed, $\phi_{\rm v} \approx 0$, the model atmosphere must have a $\Delta \phi_{tot} \sim 0.2$. 
For models with lower $\Delta \phi_{tot}$ the line splitting occurs too early in the pulsation cycle.

This is important to note when modelling lines that vary due to shock propagation through the stellar layers.
If correct phase information is to be retrieved, either models with $\phi_{tot} \sim 0.2$ need to be used, or a phase shift of $\Delta \phi_{\rm v} \sim 0.2$ (corresponding to need to a difference of $\sim 0.3$ if comparing $\phi_{\rm v}$ with $\phi_{bol}$) needs to be applied. 
The results also indicate that real Mira stars do have an inherent phase shift between the ascending luminosity and the radial movement of the surface.

\section{Comparison to 1D pulsation models}

\begin{figure*}
\centering
   \includegraphics[width=\hsize]{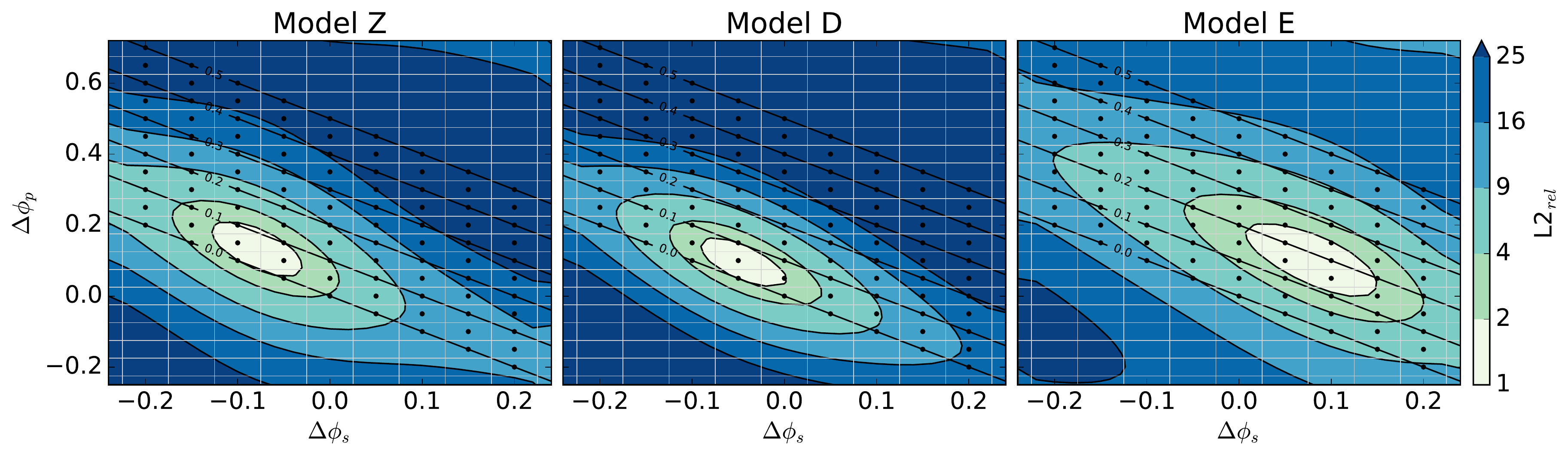}
      \caption{Comparison between the 1D pulsation models, from \citet{bessell_phase_1996}, and the boundary conditions described in Sect. \ref{ib} and Appendix\,\ref{app1}, defined by $\Delta \phi_s$ and $\Delta \phi_p$. The colours show the relative least-squares error (L2$_{rel}$), and the points indicates our calculated models and lines show $\Delta \phi_{tot} = \Delta \phi_s+\Delta \phi_p $.  \textbf{Left and middle:} Model Z and model D from \citet{bessell_phase_1996}, which are fundamental mode pulsator models of o Ceti.  \textbf{Right:} Model E from \citet{bessell_phase_1996}, a first-overtone pulsator model of o Ceti.}
         \label{bes_comp}
   \end{figure*}

The results found in this work are compared with theoretical predictions for the pulsation properties of AGB stars. 
Self-excited pulsation models of AGBs can be divided into two categories: 1D models and 3D models. 
One-dimensional self-excited radial pulsation models have been developed and worked on for decades \citep[see e.g.][]{wood_models_1974, tuchman_miras_1979,fox_theoretical_1982, ostlie_linear_1986,wood_pulsation_1990}, and contemporary models include non-linear effects and turbulent pressure \citep[e.g.][]{ireland_dynamical_2008, ireland_dynamical_2011,wood_pulsation_2015}. 
The drawback, however, is the treatment of convective energy transport, which is simulated with mixing length theory, and is mentioned as the major shortcoming in this generation of models \citep[see e.g.][ for a detailed discussion]{ barthes_pulsation_1998}. 
Three-dimensional interior pulsation models with realistic hydro-dynamical treatment should provide a more realistic description of convection, and therefore the pulsation properties; however, these models are still in their infancy, and only a small part of the relevant stellar parameter space explored so far \citep[see][]{freytag_three-dimensional_2008,freytag_global_2017}. Furthermore, possible 3D effects on mass loss have not yet been investigated.

Here we compare the different inner boundary conditions explored to results from 1D self-excited models, described in \citet{bessell_phase_1996} where non-linear models representing the prototypical Mira, o Ceti, were calculated and explicit descriptions of the variations in the sub-photospheric layers are given. 
Two fundamental mode models (model Z and D) and one first overtone model (model E) were examined. However the results, and also later investigations, suggest that Mira variables are fundamental mode pulsators. 
The first overtone model did not achieve appropriate luminosity amplitude and a factor of six had to be applied in \citet{bessell_phase_1996}. So while comparisons to all three models are included here, the first-overtone model (model E) is probably not a realistic representation of Mira stars.

\citet{bessell_phase_1996} provide Fourier coefficients to the time dependence of the luminosity and radius for the three models for the region with mean temperature of $\sim 4000-5000K$.
This is close to the location of the inner boundary of the DARWIN models. 
These variations of the photospheric layers from the 1D pulsation models are therefore directly comparable to the inner boundary condition for the DARWIN models. 
A comparison is made between different DARWIN inner boundary conditions and the variations given by \citet{bessell_phase_1996}, which are rescaled to have the same amplitude and period. This comparison is evaluated by using the a relative least-squares error (LSE, the L2 norm), which is the mean quadratic difference at each point between the curves, as seen in Eq.\,\ref{l2}, then normalised to the smallest value. The result for each model presented in \citet{bessell_phase_1996} can be seen in Fig.\,\ref{bes_comp}.

The boundary conditions most similar to the two fundamental mode pulsators given by \citet{bessell_phase_1996} are slightly asymmetric (with $\Delta \phi_s \sim -0.05$) with an offset (with $\Delta \phi_p \sim 0.1$), resulting in a total phase shift $\Delta \phi_{tot} \sim 0.05$. 
This is a smaller phase shift than found by comparing the DARWIN models with different boundary conditions and RV curves in Sect. \ref{srv_comp}. 
It overlaps with DARWIN models that produce realistic wind velocities and mass-loss rates, however. 
The first overtone pulsator model from \citet{bessell_phase_1996} predicts asymmetry (with $\Delta \phi_s \sim 0.1$) and with an offset (with $\Delta \phi_p \sim 0.1$), resulting in $\Delta \phi_{tot} \sim 0.2$. This model is  discarded as unrealistic in the \citet{bessell_phase_1996}. 

%
%

\section{Discussion}

Under the assumption that the results for model M2, representing a prototypical Mira, can be generalised to other Mira stars we can draw conclusions about which inner boundary condition agrees best with observations and about what the consequences are of using a generic boundary condition, like that of the standard DARWIN models, for the resulting atmospheres.

In our original sample of 99 models with a broad range of inner boundary conditions, 29 models reproduced realistic combinations of mass-loss rates and wind velocities. There were some systematic effects; $\Delta \phi_{tot}$ ranged between 0 and 0.3, and most of the models had a slightly asymmetric shape with $\Delta \phi_{s} < 0.0$. This sample could be reduced further, by comparing the model RV curves with compilations of CO $\Delta v = 3$ RV curves for Mira stars.

The RV comparison indicates that a $\Delta \phi_{tot}\sim 0.2$ is needed for the splitting of this line to occur at correct $\phi_{\rm v}$. As no further information about the shape can be derived from this comparison, any model with $\Delta \phi_{tot}\sim 0.2$ should reproduce line splitting at the correct $ \phi_{\rm v}$.

The $(V-K)$ vs $(J-K)$ colour-colour loops were also studied for the 29 models that produced satisfying mass-loss rates and wind velocities.  
While there was a spread in the ellipse properties, it was comparable to the same size as the spread in observed colour-colour loops. 
This is not always the case as models with unrealistic temperature structures and resulting molecule abundances in turn result in results in unrealistic loops, a fact that was explored in \cite{bladh_exploring_2013, bladh_exploring_2015}.

The similarities of the model loops here is probably a selection effect as photometry calculations are only performed on models with similar mass-loss rates and velocity properties. 
These models have similar atmospheric structures, which in turn result in similar colour-colour loops. 
Because no models were obviously unrealistic, no models were discarded using the colour-colour loop criterion.

The combined conclusion when looking at both wind velocity and mass-loss rate, and that of the RV curve comparison, the ideal inner boundary condition has a $\Delta \phi_s$ between -0.2, and 0.0, with $\Delta \phi_{tot}\sim 0.2$. 
There is thus a degeneracy when it comes to the shape of the luminosity variation.

It is not possible to pin down one ideal inner boundary condition from these tests, only that it should have $\Delta \phi_{tot}\sim 0.2$. 
However, the results do indicate that there is an inherent offset between the radial movement and the luminosity variation of Mira stars.

The standard boundary condition, which has previously been used extensively in grid calculations such as \citet{eriksson_synthetic_2014} and \citet{bladh_exploring_2015}, produces realistic wind velocities and mass-loss rate combinations for the M2 model.
If phase information about the ascension of the shock wave is not of interest, the standard boundary condition can therefore be used in the models.

Rescaling the photospheric variations of the luminosity and the gas layers found by 1D radial pulsation models from \cite{bessell_phase_1996}, and comparing them with the inner boundary conditions used in this work we find an overlap with slightly asymmetric DARWIN models ($\Delta \phi_s \sim -0.05$) with a small $\Delta \phi_{tot}$ of 0.05. 
The DARWIN models with these inner boundary conditions do reproduce realistic combination of wind velocities and mass-loss rates and good colour-colour loops; however, the line splitting in the RV curves occurs too early in the pulsation cycle when compared to observed RV curves. 

%
%

\section{Summary and conclusions}

In this paper we investigate the influence of the inner boundary conditions, used in the DARWIN models to simulate the observable variations of AGB stars, to evaluate the effects on the resulting atmosphere structure and wind properties.

The results can be summarised as follows:

\begin{itemize}
\item The DARWIN models are sensitive to the inner boundary, and using different boundary conditions can result in significant differences for the mass-loss rates (about three orders of magnitude) and wind velocities (about one order of magnitude). However, not all of the resulting models are realistic; 
\item The mass-loss rates seem to saturate (here at $1.5 \times 10^{-6}$ $M_{\odot}$ yr$^{-1}$), but changing the inner boundary will change the wind velocity;
\item All the models with boundary conditions such that they produced realistic combinations of mass-loss rates and wind velocities, also resulted in synthetic colour-colour loops that agree with observations; 
\item There is an inherent phase difference between $\phi_{bol}$ and $\phi_{\rm v}$ for the DARWIN models of $\sim 0.1$. This has previously been suggested to be the case for observed stars;
\item Phase information can be derived from RV curves; DARWIN models with a $\Delta \phi_{tot} \sim 0.2$ results in a line splitting phase that corresponds to observations;
\item The 1D radial pulsation models of \cite{bessell_phase_1996} indicate that the total phase difference $\Delta \phi_{tot}$ is smaller than $\Delta \phi_{tot} \sim 0.2$;
\item Using the standard boundary condition results in realistic mass-loss rates, wind velocities and colour-colour loops in the case of M-stars; however, the line splitting of the CO dv=3 line will occur at the wrong $\phi_{\rm v}$. 
\end{itemize}

It can therefore be concluded that grid studies of DARWIN models, such as \cite{eriksson_synthetic_2014} and \cite{bladh_exploring_2015}, that use the standard inner boundary condition should produce realistic mass-loss rates and wind expansion velocities.

\begin{acknowledgements}
The observational results (RV data as well as FTS spectra of $\chi$ Cyg, S Cep, and W Hya) were kindly provided by Th. Lebzelter and K. Hinkle.
S.H. would like to acknowledge the support from the Swedish Research Council (Vetenskapsr{\aa}det). 
The computations were performed on resources provided by the Swedish National Infrastructure for Computing (SNIC) at UPPMAX. 
\end{acknowledgements}

\bibliographystyle{aa} 
\bibliography{ref,ref2} 

\begin{thebibliography}{45} \expandafter\ifx\csname natexlab
\endcsname\relax\def\natexlab#1{#1}\fi 

\bibitem[{Aringer {et~al.}(2016)Aringer, Girardi, Nowotny, Marigo, \& Bressan}]{aringer_synthetic_2016} Aringer, B., Girardi, L., Nowotny, W., Marigo, P., \& Bressan, A. 2016, MNRAS, 457, 3611 

\bibitem[{Barthes(1998)}]{barthes_pulsation_1998} Barthes, D. 1998, A\&A, 333, 647 

\bibitem[{Bessell(1990)}]{bessell_ubvri_1990} Bessell, M.~S. 1990, PASP, 102, 1181 

\bibitem[{Bessell \& Brett(1988)}]{bessell_jhklm_1988} Bessell, M.~S. \& Brett, J.~M. 1988, PASP, 100, 1134 

\bibitem[{Bessell {et~al.}(1996)Bessell, Scholz, \& Wood}]{bessell_phase_1996} Bessell, M.~S., Scholz, M., \& Wood, P.~R. 1996, A\&A, 307, 481 

\bibitem[{Bladh {et~al.}(2015)Bladh, H{\"o}fner, Aringer, \& Eriksson}]{bladh_exploring_2015} Bladh, S., H{\"o}fner, S., Aringer, B., \& Eriksson, K. 2015, A\&A, 575, A105 

\bibitem[{Bladh {et~al.}(2013)Bladh, H{\"o}fner, Nowotny, Aringer, \& Eriksson}]{bladh_exploring_2013} Bladh, S., H{\"o}fner, S., Nowotny, W., Aringer, B., \& Eriksson, K. 2013, A\&A, 553, A20 

\bibitem[{Bowen(1988)}]{bowen_dynamical_1988} Bowen, G.~H. 1988, ApJ, 329, 299 

\bibitem[{Chandler {et~al.}(2007)Chandler, Tatebe, Wishnow, Hale, \& Townes}]{chandler_asymmetries_2007} Chandler, A.~A., Tatebe, K., Wishnow, E.~H., Hale, D. D.~S., \& Townes, C.~H. 2007, ApJ, 670, 1347 

\bibitem[{Eriksson {et~al.}(2014)Eriksson, Nowotny, H{\"o}fner, Aringer, \& Wachter}]{eriksson_synthetic_2014} Eriksson, K., Nowotny, W., H{\"o}fner, S., Aringer, B., \& Wachter, A. 2014, A\&A, 566, A95 

\bibitem[{Fox \& Wood(1982)}]{fox_theoretical_1982} Fox, M.~W. \& Wood, P.~R. 1982, ApJ, 259, 198 

\bibitem[{Freytag \& H{\"o}fner(2008)}]{freytag_three-dimensional_2008} Freytag, B. \& H{\"o}fner, S. 2008, A\&A, 483, 571 

\bibitem[{Freytag {et~al.}(2017)Freytag, Liljegren, \& H{\"o}fner}]{freytag_global_2017} Freytag, B., Liljegren, S., \& H{\"o}fner, S. 2017, A\&A, 600, A137 

\bibitem[{Gail {et~al.}(2016)Gail, Scholz, \& Pucci}]{gail_silicate_2016} Gail, H.-P., Scholz, M., \& Pucci, A. 2016, A\&A, 591, A17 

\bibitem[{Gail \& Sedlmayr(1999)}]{gail_mineral_1999} Gail, H.-P. \& Sedlmayr, E. 1999, A\&A, 347, 594 

\bibitem[{Gautschy-Loidl {et~al.}(2004)Gautschy-Loidl, H{\"o}fner, J{\o}rgensen, \& Hron}]{gautschy-loidl_dynamic_2004} Gautschy-Loidl, R., H{\"o}fner, S., J{\o}rgensen, U.~G., \& Hron, J. 2004, A\&A, 422, 289 

\bibitem[{Gobrecht {et~al.}(2016)Gobrecht, Cherchneff, Sarangi, Plane, \& Bromley}]{gobrecht_dust_2016} Gobrecht, D., Cherchneff, I., Sarangi, A., Plane, J. M.~C., \& Bromley, S.~T. 2016, A\&A, 585, A6 

\bibitem[{Gonz{\'a}lez~Delgado {et~al.}(2003)Gonz{\'a}lez~Delgado, Olofsson, Kerschbaum, Sch{\"o}ier, Lindqvist, \& Groenewegen}]{gonzalez_delgado_``thermal_2003} Gonz{\'a}lez~Delgado, D., Olofsson, H., Kerschbaum, F., {et~al.} 2003, A\&A, 411, 123 

\bibitem[{Hinkle {et~al.}(1982)Hinkle, Hall, \& Ridgway}]{hinkle_time_1982} Hinkle, K.~H., Hall, D. N.~B., \& Ridgway, S.~T. 1982, ApJ, 252, 697 

\bibitem[{H{\"o}fner(2008)}]{hofner_winds_2008} H{\"o}fner, S. 2008, A\&A, 491, L1 

\bibitem[{{H{\"o}fner}(2015)}]{hof15} {H{\"o}fner}, S. 2015, in Astronomical Society of the Pacific Conference Series, Vol. 497, Why Galaxies Care about AGB Stars III: A Closer Look in Space and Time, ed. F.~{Kerschbaum}, R.~F. {Wing}, \& J.~{Hron}, 333 

\bibitem[{H{\"o}fner {et~al.}(2016)H{\"o}fner, Bladh, Aringer, \& Ahuja}]{hofner_dynamic_2016-1} H{\"o}fner, S., Bladh, S., Aringer, B., \& Ahuja, R. 2016, A\&A, 594, A108 

\bibitem[{H{\"o}fner {et~al.}(2003)H{\"o}fner, Gautschy-Loidl, Aringer, \& J{\o}rgensen}]{hofner_dynamic_2003} H{\"o}fner, S., Gautschy-Loidl, R., Aringer, B., \& J{\o}rgensen, U.~G. 2003, A\&A, 399, 589 

\bibitem[{Ireland {et~al.}(2008)Ireland, Scholz, \& Wood}]{ireland_dynamical_2008} Ireland, M.~J., Scholz, M., \& Wood, P.~R. 2008, MNRAS, 391, 1994 

\bibitem[{Ireland {et~al.}(2011)Ireland, Scholz, \& Wood}]{ireland_dynamical_2011} Ireland, M.~J., Scholz, M., \& Wood, P.~R. 2011, MNRAS, 418, 114 

\bibitem[{Karovicova {et~al.}(2013)Karovicova, Wittkowski, Ohnaka, Boboltz, Fossat, \& Scholz}]{karovicova_new_2013} Karovicova, I., Wittkowski, M., Ohnaka, K., {et~al.} 2013, A\&A, 560, A75 

\bibitem[{Lebzelter(2011)}]{lebzelter_shapes_2011} Lebzelter, T. 2011, AN, 332, 140 

\bibitem[{{Lebzelter} \& {Hinkle}(2002)}]{leb02} {Lebzelter}, T. \& {Hinkle}, K.~H. 2002, in Astronomical Society of the Pacific Conference Series, Vol. 259, IAU Colloq. 185: Radial and Nonradial Pulsationsn as Probes of Stellar Physics, ed. C.~{Aerts}, T.~R. {Bedding}, \& J.~{Christensen-Dalsgaard}, 556 

\bibitem[{Lebzelter {et~al.}(2001)Lebzelter, Hinkle, \& Aringer}]{lebzelter_4_2001} Lebzelter, T., Hinkle, K.~H., \& Aringer, B. 2001, A\&A, 377, 617 

\bibitem[{Liljegren {et~al.}(2016)Liljegren, H{\"o}fner, Nowotny, \& Eriksson}]{liljegren_dust-driven_2016} Liljegren, S., H{\"o}fner, S., Nowotny, W., \& Eriksson, K. 2016, A\&A, 589, A130 

\bibitem[{Norris {et~al.}(2012)Norris, Tuthill, Ireland, Lacour, Zijlstra, Lykou, Evans, Stewart, \& Bedding}]{norris_close_2012} Norris, B. R.~M., Tuthill, P.~G., Ireland, M.~J., {et~al.} 2012, Nature, 484, 220 

\bibitem[{Nowotny {et~al.}(2005{\natexlab{a}})Nowotny, Aringer, H{\"o}fner, Gautschy-Loidl, \& Windsteig}]{nowotny_atmospheric_2005-1} Nowotny, W., Aringer, B., H{\"o}fner, S., Gautschy-Loidl, R., \& Windsteig, W. 2005{\natexlab{a}}, A\&A, 437, 273 

\bibitem[{Nowotny {et~al.}(2011)Nowotny, Aringer, H{\"o}fner, \& Lederer}]{nowotny_synthetic_2011} Nowotny, W., Aringer, B., H{\"o}fner, S., \& Lederer, M.~T. 2011, A\&A, 529, A129 

\bibitem[{Nowotny {et~al.}(2010)Nowotny, H{\"o}fner, \& Aringer}]{nowotny_line_2010} Nowotny, W., H{\"o}fner, S., \& Aringer, B. 2010, A\&A, 514, A35 

\bibitem[{Nowotny {et~al.}(2005{\natexlab{b}})Nowotny, Lebzelter, Hron, \& H{\"o}fner}]{nowotny_atmospheric_2005} Nowotny, W., Lebzelter, T., Hron, J., \& H{\"o}fner, S. 2005{\natexlab{b}}, A\&A, 437, 285 

\bibitem[{Ohnaka {et~al.}(2012)Ohnaka, Hofmann, Schertl, Weigelt, Malbet, Massi, Meilland, \& Stee}]{ohnaka_spatially_2012} Ohnaka, K., Hofmann, K.-H., Schertl, D., {et~al.} 2012, A\&A, 537, A53 

\bibitem[{Ohnaka {et~al.}(2016)Ohnaka, Weigelt, \& Hofmann}]{ohnaka_clumpy_2016} Ohnaka, K., Weigelt, G., \& Hofmann, K.-H. 2016, A\&A, 589, A91 

\bibitem[{Olofsson {et~al.}(2002)Olofsson, Gonz{\'a}lez~Delgado, Kerschbaum, \& Sch{\"o}ier}]{olofsson_mass_2002} Olofsson, H., Gonz{\'a}lez~Delgado, D., Kerschbaum, F., \& Sch{\"o}ier, F.~L. 2002, A\&A, 391, 1053 

\bibitem[{Ostlie \& Cox(1986)}]{ostlie_linear_1986} Ostlie, D.~A. \& Cox, A.~N. 1986, ApJ, 311, 864 

\bibitem[{Ramstedt {et~al.}(2008)Ramstedt, Sch{\"o}ier, Olofsson, \& Lundgren}]{ramstedt_reliability_2008} Ramstedt, S., Sch{\"o}ier, F.~L., Olofsson, H., \& Lundgren, A.~A. 2008, A\&A, 487, 645 

\bibitem[{Tuchman {et~al.}(1979)Tuchman, Sack, \& Barkat}]{tuchman_miras_1979} Tuchman, Y., Sack, N., \& Barkat, Z. 1979, ApJ, 234, 217 

\bibitem[{Winters {et~al.}(2000)Winters, Le~Bertre, Jeong, Helling, \& Sedlmayr}]{winters_systematic_2000} Winters, J.~M., Le~Bertre, T., Jeong, K.~S., Helling, C., \& Sedlmayr, E. 2000, A\&A, 361, 641 

\bibitem[{Wood(1974)}]{wood_models_1974} Wood, P.~R. 1974, ApJ, 190, 609 

\bibitem[{Wood(1990)}]{wood_pulsation_1990} Wood, P.~R. 1990, in From {{Miras}} to {{Planetary Nebulae}}: {{Which Path}} for {{Stellar Evolution}}?, 67--84 

\bibitem[{Wood(2015)}]{wood_pulsation_2015} Wood, P.~R. 2015, MNRAS, 448, 3829 
\end{thebibliography}

\begin{appendix} 

\section{Description of the boundary condition}
\label{app1}

\subsection{Inner boundary}

Originally in the DARWIN code the effects of stellar pulsation were described by two variable physical quantities at the inner boundary: $R_{in}(t)$ and $L_{in}(t)$. 
The first $R_{in}(t)$, is the variation in the innermost gas layer. 
This lower boundary layer is impermeable, meaning no gas flows across it. 
The second, $L_{in}(t)$ describes the variation in the luminosity. 
The inner boundary is places below the stellar photosphere, at around $R \approx 0.9 R_\star$. 
The $R_{in}(t)$ variation has the form

\begin{equation}
\label{eqn1}
R_{in}(t) = R_0 + \frac{\Delta u_p P}{2 \pi} \sin{\left ( \frac{2 \pi }{P} t \right)}
\end{equation}

\noindent
where $\Delta u_p$ is the velocity amplitude, P is the pulsation period, and $R_0$ the average radial distance of the boundary. This corresponds to a gas velocity of 

\begin{equation}
\label{eqn11}
u_{in}(t) = \Delta u_p \cos\left ( \frac{2 \pi }{P} t \right)
\end{equation}

\noindent
Assuming a constant flux at the inner boundary the luminosity becomes $L_{in}(t) \propto R^2_{in}(t)$. To better match observations of the bolometric flux variation a free parameter $f_L$ was introduced in \citet{gautschy-loidl_dynamic_2004}, so that the amplitude of luminosity can be adjusted separately from the radial amplitude. The original form of the luminosity is

\begin{align}
\label{eqn2}
\Delta L_{in}(t) &= L_{in} - L_0= f_L \left (\frac{R^2_{in}(t) - R^2_0}{R^2_0} \right ) \times L_0 \nonumber  \\ 
& =  f_L \left ( \left[1 + \frac{\Delta u_p P}{R_0 2 \pi} \sin{\left ( \frac{2 \pi }{P} t \right)}\right]^2- 1 \right ) \times L_0
\end{align}

\noindent
where $L_0$ is the average luminosity of the star. 
In this paper we alter the luminosity variation, but keep the form of the radial variation. 

If a phase shift $\Delta \phi_p$ is introduced between $R_{in}(t)$ and $L_{in}(t)$, but the shape is kept unchanged, the luminosity variation becomes

\begin{equation}
\label{eqn3}
\Delta L_{in}(t, \Delta \phi_p) =  f_L \left (\left[1 + \frac{\Delta u_p P}{2 \pi R_0} \sin{\left ( 2 \pi \left(\frac{t}{P} + \Delta \phi_p \right)  \right)}\right]^2- 1 \right ) \times L_0
\end{equation}

\noindent

The asymmetric luminosity variation is achieved by describing the luminosity variation at the inner boundary by a modified, smoothed Fourier saw-tooth wave curve as

\begin{equation}
\label{ib_four}
\Delta L_{in}(t, w(\Delta \phi_s), \Delta \phi_p) = k \sum^N_{n=1} \frac{1}{n^{w(\Delta \phi_s)}} \sin \left( \frac{2 \pi n}{P} \left(t + \Delta \phi_p \right)\right)
\end{equation}

\noindent
where k is the amplitude set to match the amplitude from Eq. (\ref{eqn2}), P is the pulsation period and $w(\Delta \phi_s)$ is a smoothing factor. 
If $w(\Delta \phi_s) = 1$ we retain the saw-tooth shape; however, when $w$ is increased $\Delta L_{in}$ will approach a sinusoidal curve. 
As seen in Fig.\,\ref{bc_ex}, the $\Delta \phi _s$ is a measure of how asymmetric the curve is, specifying how far from the original $L_{in}$ the phase of the maximum has been shifted. 
This measure is analogous to $\Delta \phi_p$. 

This form also allows for different values of $\Delta \phi_p$.

\subsection{Estimation of w}

\begin{figure}
\centering
\includegraphics[width=\hsize]{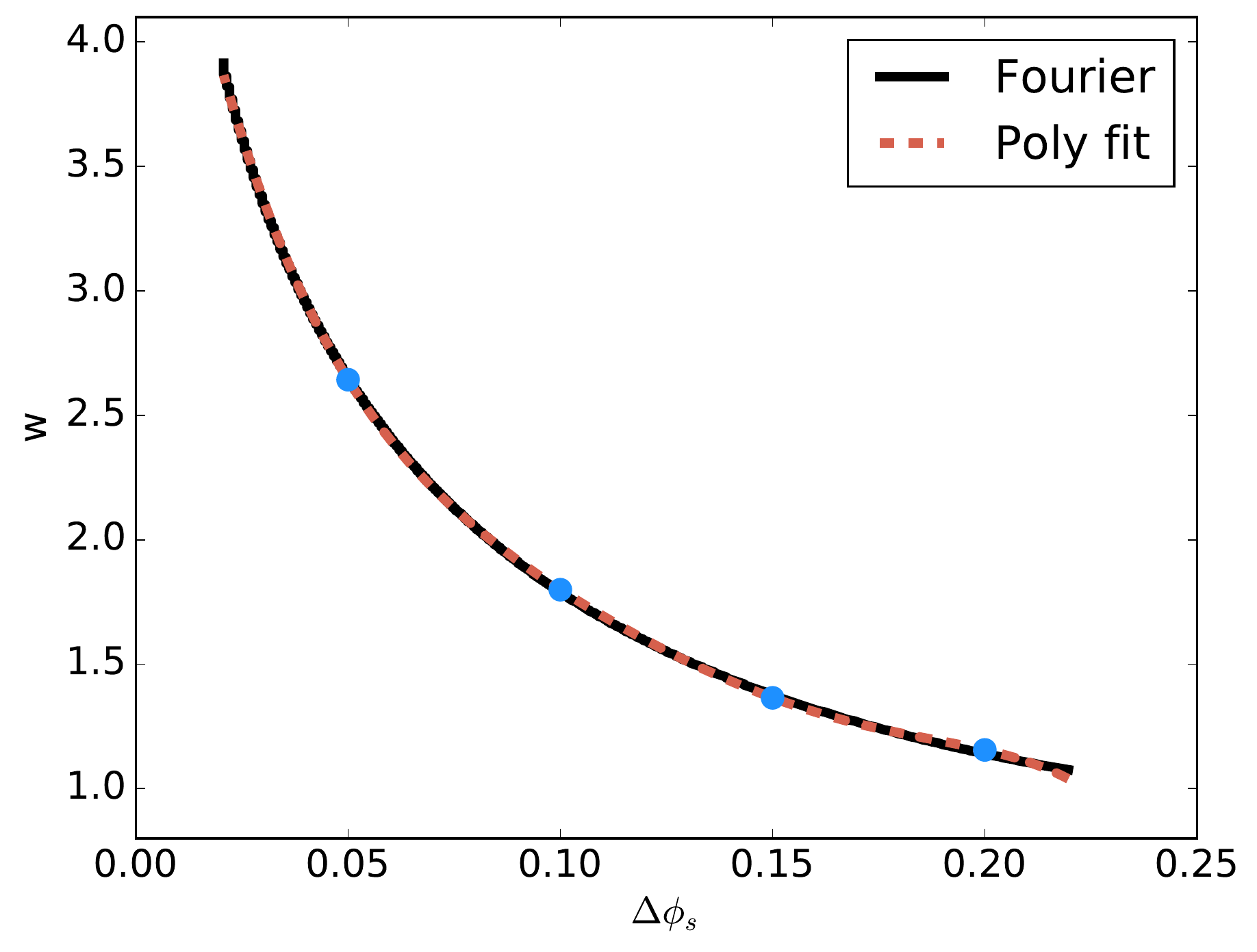}
     \caption{Exact $w = f(\Delta \phi_s)$, from the Fourier series Eq. \ref{ib_four}, calculated with 10000 terms and the polynomial fit from Eq. \ref{wds}, using the terms from Table \ref{phis-w-tab}. The blue circles represent the values of $\Delta \phi_s$ used in this work. }
        \label{phis-w}
\end{figure}

\begin{table}[t]
\centering
\caption{Polynomial constants, for Eq. \ref{wds}.}
\label{phis-w-tab}
\begin{tabular}{c c c} 
\hline\hline 
m & $a_m$      & $\Delta \phi_s^{ m}$ \\ 
\hline \\
0 &-5.3309e+04    & $\Delta \phi_s^0$ \\
1 & 3.6367e+04  & $\Delta \phi_s^1$ \\
2 & -9.7661e+03  & $\Delta \phi_s^2$ \\
3 &  1.3385e+03  & $\Delta \phi_s^3$ \\
4 & -1.0460e+02    & $\Delta \phi_s^4$ \\
5 &  5.5368e+00 & $\Delta \phi_s^5$\\
\hline
\end{tabular}
\end{table}

The parameter $w$, from Eq. (\ref{ib_four}), is a function of $\Delta \phi_s$ and the relationship between $w$ and $\Delta \phi_s$ can be seen in Fig.\,\ref{phis-w}. For a simpler extraction of the $w$-value a polynomial is fitted, of the form

\begin{equation}
\label{wds}
w(\Delta \phi_s) \approx \sum^M_{m=0} a_m \times \Delta \phi_s^m
\end{equation}

A good fit was found for M = 5. Table  \ref{phis-w-tab} shows the constants for this fit.

\subsection{Number of terms used}
\label{a3}

   \begin{figure}
   \centering
   \includegraphics[width=\hsize]{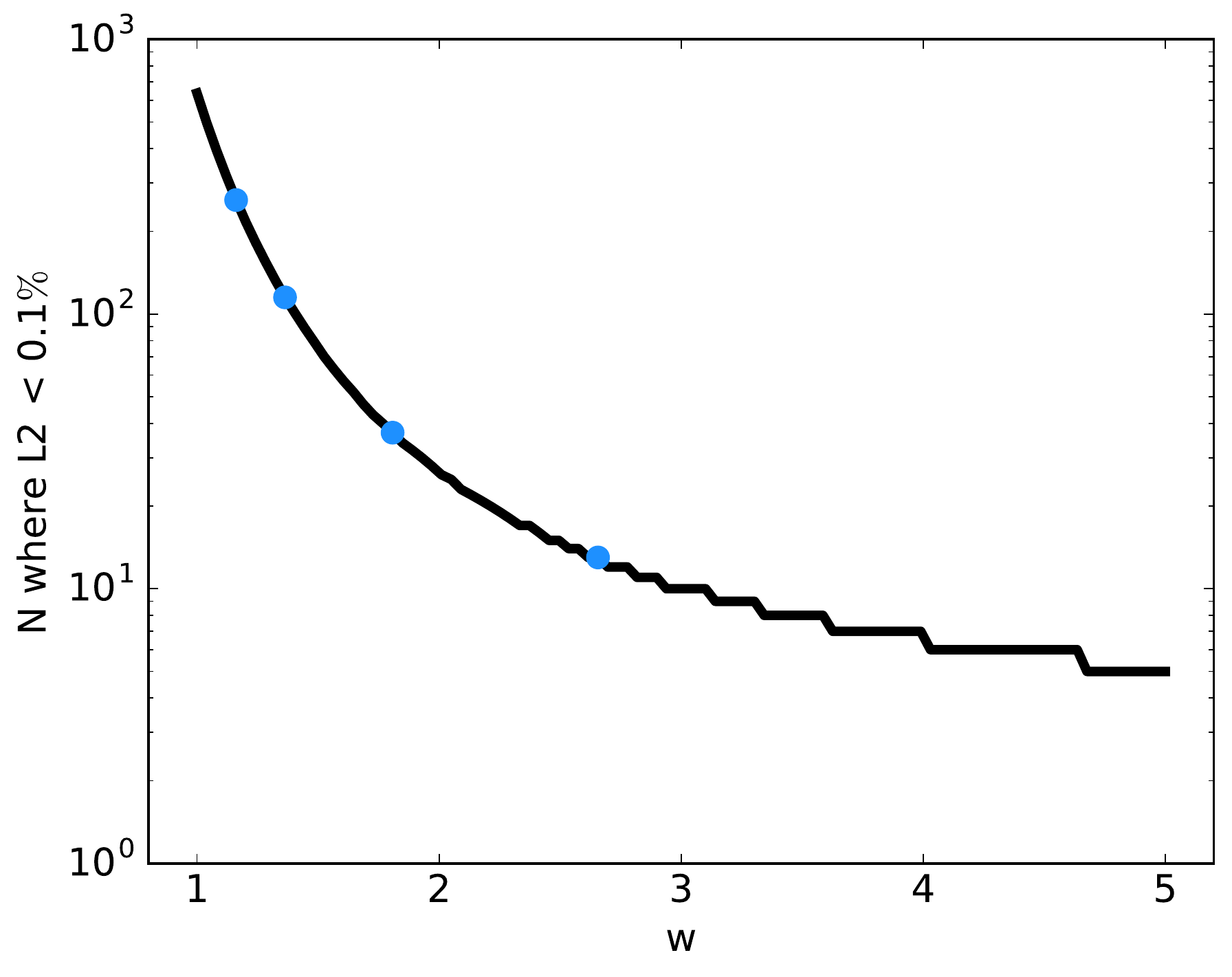}
      \caption{Number of Fourier terms to reach an L2 value of less than 0.1$\%$.}
         \label{N-w}
   \end{figure}

When using the Fourier series from Eq. (\ref{ib_four}), the number of terms $N$ needed to avoid significant unwanted overshooting close to the discontinuity (Gibbs phenomenon) depends on the value of $\Delta \phi_s$. With a higher $\Delta \phi_s$, the luminosity variation will be more asymmetric and more prone to overshooting, which requires a larger $N$. 

To estimate the $N$ number of Fourier terms needed for convergence at different $\Delta \phi_s$, $\Delta L_{in}$ for one period calculated using $N$ Fourier terms is compared to one period with $N+1$ terms.  
When the difference between $\Delta L_{in}(N)$ and $\Delta L_{in}(N+1)$ becomes small, it is assumed that $\Delta L_{in}(N)$ has converged. 

The difference between the two curves is measured with the least-squares error, defined as
\begin{equation}
\label{l2}
L2 = \sqrt{ \frac{1}{x_{steps}} \sum^{x_{steps}} _i  \left( f_i -g_i \right)^2}
\end{equation}
with $f_i = \Delta L_{in}(N+1)_i$ and $g_i = \Delta L_{in}(N)_i$. This is a measure of the the average difference between two points on the curves, and is usually denoted as the least-squares error or the L2 norm. Convergence is assumed when $L2 < 0.001$.

An overview of the number of Fourier terms $N$ used in this work, for each value of $\Delta \phi_s$, is seen in Table \ref{N-w-tab} and Fig. \ref{N-w}.
%
%
%
%

%
%

\begin{table}
\centering
\caption{Number of terms used in this work, for different values of $\Delta \phi_s$.}
\label{N-w-tab}
\begin{tabular}{c c c} 
\hline\hline 
$\Delta \phi_s$ & $w$      &  N where $L2 <0.001 $ \\ 
\hline \\
0.05 & 2.66 &  13 \\
0.1 &1.81  & 37 \\
0.15 & 1.36 & 115 \\
0.2 & 1.16  &  260 \\
\hline
\end{tabular}
\end{table}
   
\end{appendix}

\end{document}